\documentclass[aps,prd,preprint,groupedaddress,eqsecnum]{revtex4}
\usepackage[xdvi]{graphicx}
\newcommand{\labell}[1]{\label{#1}}
\newcommand{\bbibitem}[1]{\bibitem{#1}}

\newcommand{\be}{\begin{equation}}
\newcommand{\ee}{\end{equation}}
\newcommand{\bea}{\begin{eqnarray}}
\newcommand{\eea}{\end{eqnarray}}

\begin{document}

%\preprint{}

\preprint{hep-th/0405048, UPR-T-1079}

\title{Beyond the veil: Inner horizon instability and holography}


\author{Vijay Balasubramanian}
\email{vijay@endive.hep.upenn.edu}
%\thanks{}
%\affiliation{}
\author{Thomas S.\ Levi}
\email{tslevi@student.physics.upenn.edu} \affiliation{David Rittenhouse Laboratories, University of Pennsylvania, Philadelphia, PA\quad
19104-6396}


\date{\today}

\begin{abstract}
We show that scalar perturbations  of the eternal, rotating BTZ black hole should lead to an instability of the inner (Cauchy) horizon,
preserving strong cosmic censorship.   Because of backscattering from the geometry,  plane wave modes have a divergent stress tensor at the
event horizon, but suitable wavepackets avoid this difficulty, and are dominated at late times by quasinormal behavior.  The wavepackets
have cuts in the complexified coordinate plane that are controlled by requirements of continuity, single-valuedness and positive energy.
Due to a focusing effect,  regular wavepackets nevertheless have a divergent stress-energy at the inner horizon, signaling an instability.
This instability, which is localized behind the event horizon, is detected holographically as a breakdown in the semiclassical computation
of dual CFT expectation values in which the analytic behavior of wavepackets in the complexified coordinate plane plays an integral role.
In the dual field theory, this is interpreted as an encoding of physics behind the horizon in the entanglement between otherwise independent
CFTs.
\end{abstract}

%\keywords{}

\maketitle \tableofcontents

\section{Introduction}

Behind the event horizon of rotating black holes lies an inner horizon  which  shields a region  containing naked timelike singularities. In
asymptotically flat space, numerical simulations and analytic studies have shown that a generic, small perturbation can cause such an inner
horizon to itself collapse to a singularity \cite{numerical instability, flat space blueshifts} (for a more modern study in the context of
classical gravity see \cite{dafermos}). This is a manifestation of strong cosmic censorship \cite{scc}. If such an instability also exists
for black holes in AdS space, it presents an opportunity to study how a phenomenon that is entirely localized behind an event horizon is
represented holographically in a dual field theory.   To this end, we show how a generic scalar perturbation of the rotating BTZ black hole
leads to a divergent stress tensor on the inner horizon and explain how the dual CFT would detect the resulting instability.

An important subtlety is that the standard basis of plane wave modes in a BTZ black hole has a divergent stress tensor on the event horizon
due to interaction between outgoing and ingoing parts of the solution.  This necessitates the construction of wavepackets, which have cuts
in the complexified coordinate plane that are controlled by requirements of single-valuedness, continuity and positive energy.   The cuts in
the complex time plane enable an observer who only integrates over the region outside the black hole, and parts of the complexified
coordinate plane, to sense the instability at the inner horizon.  From the perspective of the dual CFT this is evidence that entanglement of
the two otherwise independent components of the dual field theory encodes the region behind the horizon.

We are familiar with the use of the complexified energy plane as a powerful tool in analyzing scattering amplitudes.  For example, a pole at
a complex energy indicates a resonance or meta-stable excitation in a system.  We are less comfortable with the complexified coordinate
plane, but there is increasing evidence that structures in complex time and space contain important information about physics in time
dependent backgrounds.  In \cite{Hemming:2002kd,kos,leviross,Fidkowski:2003nf, Kaplan:2004qe} various signatures of physics behind a black
hole horizon were associated with the analytic structure of physical quantities in complexified time.    In \cite{Balasubramanian:2003kq}
degenerations of the metric of an AdS orbifold in the complex plane of a spatial coordinate were found, surprisingly, to control unitarity
and quantization conditions in the real spacetime.  A general lesson from all these works and from the present paper might well be that in
time-dependent universes, particularly when analytic continuation to a Euclidean section is not possible, structures in the complexfied
spacetime are important for physics.

\section{The rotating BTZ black hole and its embedding}
\label{sec-btz}
\subsection{The geometry and causal structure}
The rotating BTZ black hole is the unique rotating black hole solution in three dimensions with a negative cosmological constant.  It is
locally $AdS_3$ and can be described by the metric \cite{btz,bthz}
\begin{eqnarray}
ds^2 &=& -\frac{(r^2-r_+^2)(r^2-r_-^2)}{\Lambda ^2 r^2} dt^2 + \frac{\Lambda ^2 r^2}{(r^2-r_+^2)(r^2-r_-^2)}dr^2+r^2 \Bigl( d\bar{\phi} -
\frac{r_+
r_-}{\Lambda r^2} dt \Bigr)^2  , \nonumber \\
\bar{\phi} &\sim& \bar{\phi} +2\pi \labell{btzmetric}
\end{eqnarray}
where $r_\pm$ are the locations of the outer and inner horizons respectively.   The mass and angular momentum are
\begin{equation}
M= \frac{r_+^2 + r_-^2}{\Lambda ^2} ~~~~;~~~~ J= \frac{2r_+ r_-}{\Lambda} .
\end{equation}
It will be easier for discussing the near horizon physics to introduce a different angular coordinate, $\phi_+ = \bar{\phi} - \Omega_H t$,
where
\begin{equation}
\Omega_H = {r_-  \over \Lambda r_+} \labell{outerangvel}
\end{equation}
is the angular velocity of the outer horizon. This coordinate change untwists the $\bar{\phi}$ circle at the outer horizon. The metric in
these coordinates becomes
\begin{equation}
ds^2= -\frac{(r^2-r_+^2)(r^2-r_-^2)}{\Lambda ^2 r^2} dt^2 + \frac{\Lambda ^2 r^2}{(r^2-r_+^2)(r^2-r_-^2)}dr^2+r^2 \Bigl( \frac{r_-}{\Lambda
r_+ r^2} (r^2-r_+^2) dt + d \phi_+ \Bigr)^2 . \labell{untwisted btzmetric}
\end{equation}
We can view this geometry as embedded in $AdS_3$ with identifications. $AdS_3$ is the hyperboloid $T_1^2 + T_2^2 -X_1^2 -X_2^2 = \Lambda^2$
embedded in $R^{2,2}$ with metric $ds^2 = -dT_1 ^2 -dT_2 ^2 + dX_1^2 + dX_2 ^2 $. One can realize this embedding as
\begin{eqnarray}
T_1 &=&  \sqrt{u} \cosh \Bigl( \frac{r_+}{\Lambda} \phi_+ \Bigr) , \nonumber \\
T_2 &=&  \sqrt{u-1} \sinh \Bigl(\kappa_+ t - \frac{r_-}{\Lambda} \phi_+ \Bigr) , \nonumber \\
X_1 &=& \sqrt{u} \sinh \Bigl( \frac{r_+}{\Lambda} \phi_+ \Bigr), \nonumber \\
X_2 &=& \sqrt{u-1} \cosh \Bigl(\kappa_+ t - \frac{r_-}{\Lambda} \phi_+ \Bigr) ,
 \labell{embedding}
\end{eqnarray}
where
\begin{equation}
u= \frac{r^2-r_-^2}{r_+^2 - r_-^2}  ~~~~;~~~~ \kappa_{\pm} = \frac{r_+^2 - r_-^2}{\Lambda^2 r_\pm} \, .
\end{equation}
$\kappa_\pm$ are the surface gravities at the inner and outer horizon respectively.

The spacetime can be divided into several regions determined by the norm of the Killing vector $\xi = \partial_{\phi_+}$. In region 1, we
are outside the black hole's event horizon and $\xi \cdot \xi > r_+^2$. In region 2, we are between the event horizon and the Cauchy horizon
and $r_+^2> \xi \cdot \xi >r_-^2$. In region 3 we are behind the Cauchy horizon but outside the singularity, and $r_-^2 >\xi \cdot \xi >0$.
Finally, before making the identification on $\phi_+$, a region where the norm is less than zero exists (this would be region 4). Upon
making the identification on $\phi_+$ this region would contain closed timelike curves. We cutoff the spacetime at the surface $\xi \cdot
\xi =0$, treating it as a singularity, so the spacetime has no closed timelike curves.  This singularity is timelike if $r_-\neq0$ (i.e.\
the black hole has non-vanishing angular momentum) and spacelike if $r_-=0$. We will principally be interested in the finite angular
momentum case.

\begin{figure}
\begin{center}
    \includegraphics[width=0.5\textwidth]{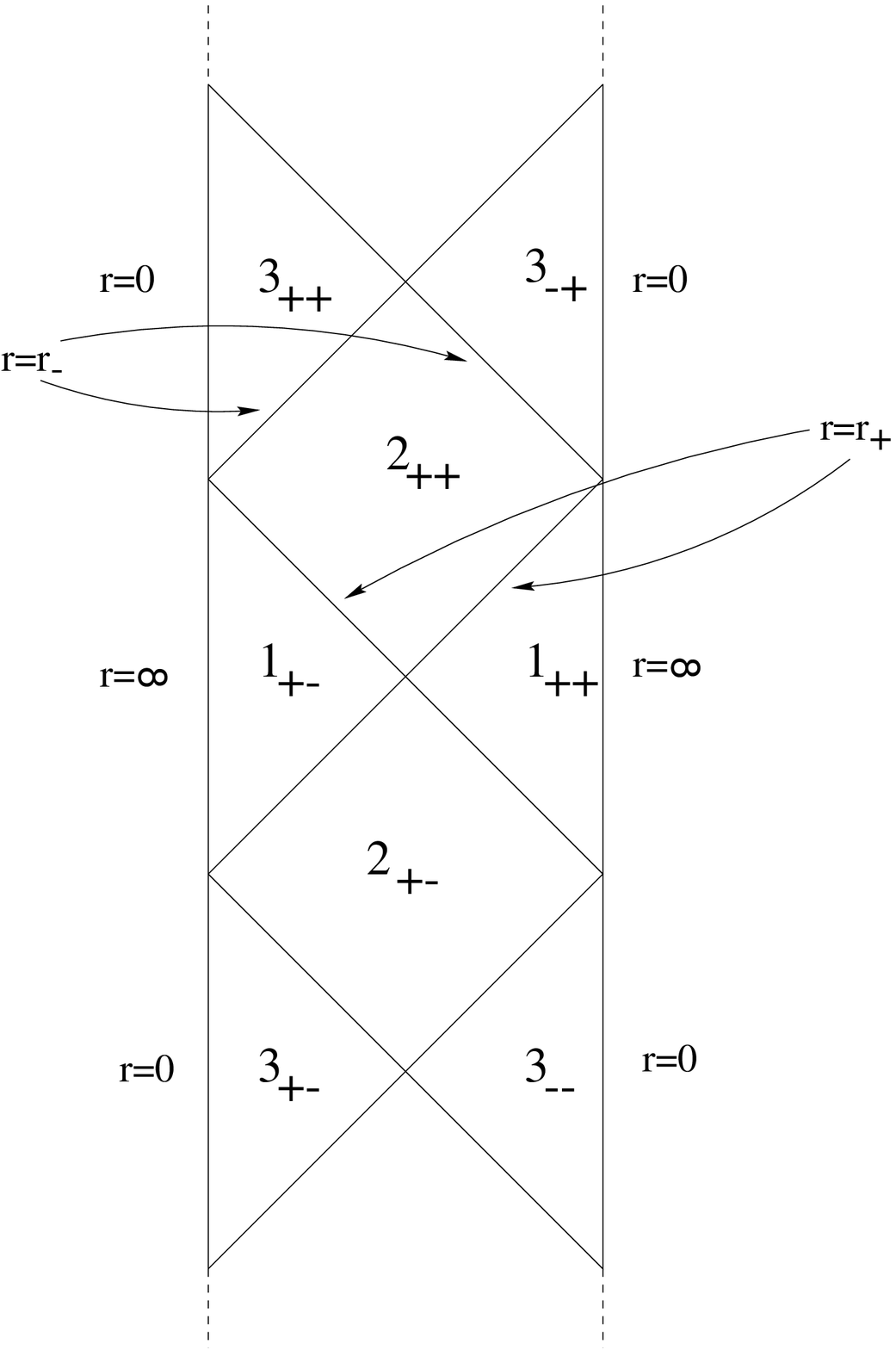}
\end{center}
\caption{Penrose diagram detailing the causal structure of the rotating BTZ black hole \cite{bthz}. The diagram repeats itself indefinitely
above and below. The vertical boundaries of the regions $3_{\pm\pm}$ are singularities while the vertical boundaries of the regions
$1_{\pm\pm}$ are asymptotically AdS boundaries.  We study an instability of the inner horizon at the locations marked $r_-$ that results
from the gravitational focusing of wavepackets. } \label{fig:pen}
\end{figure}

In terms of the embedding coordinates, $T_1^2-X_1^2 = u$ and $T_2^2 -X_2^2 = 1-u$, so that the regions can also be distinguished by
\begin{eqnarray}
\mbox{ region 1:} & T_1^2 - X_1^2 \geq 0, ~~~ &  T_2^2 - X_2^2 \leq 0 \\
\mbox{ region 2:} & T_1^2 - X_1^2 \geq 0, ~~~ & T_2^2 - X_2^2 \geq 0  \nonumber \\
\mbox{ regions 3 \& 4:} & T_1^2 - X_1^2 \leq 0, ~~~~ & T_2^2 - X_2^2 \geq 0 . \nonumber
\end{eqnarray}
Each of these regions describes several disconnected components of the maximally extended geometry which can be distinguished by the signs
of certain combinations of the embedding coordinates. Following the procedure of \cite{Hemming:2002kd,kos,leviross}, denote the different
regions by $A_{\eta_1 \eta_2}$, where $A=1,\ldots,3$ and $\eta_1,\eta_2= \pm$ are the signs of the two combinations $T_1+X_1$ and $T_2+X_2$.
The boundary of the spacetime is given by two disconnected cylinders.  Fig.\ \ref{fig:pen} is a Penrose diagram.

\begin{figure}
\begin{center}
    \includegraphics[width=0.7\textwidth]{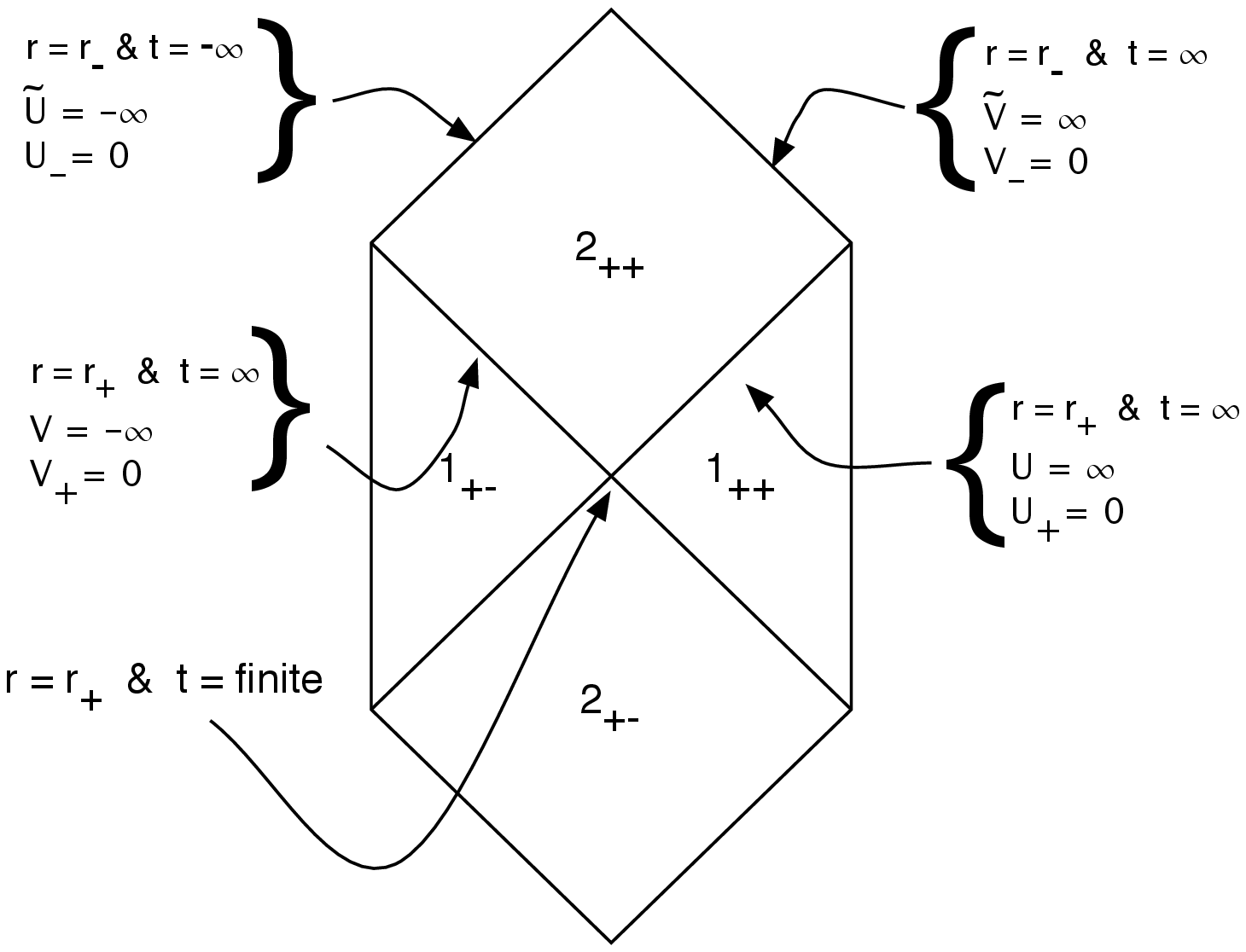}
\end{center}
\caption{The Penrose diagram for regions of interest, detailing some of the horizon structure in the various coordinate systems. The
horizons in the other regions are similar. } \label{fig:horizon}
\end{figure}

A  null geodesic starting from a location outside the black hole takes infinite BTZ coordinate time to reach either the past or future event
horizon where the coordinate system breaks down. The future (past) event horizon is line that is approached as $r \to r_+$ while $t \to
\infty$ ($t \to -\infty$).  The intersection of the past and future event horizons occurs in the limit $r \to r_+$ for any finite $t$.   (We
will define a coordinate system that makes this apparent later.) In a similar way, one finds the right (left) Cauchy horizon in the $2_{++}$
region is given by $r \to r_-$ and  $t \to + \infty$ ($t \to - \infty$).  Fig.~\ref{fig:horizon} shows the horizon structure in the $1_{++}$
and $2_{++}$ regions.

Beyond the inner horizon at $r_-$, the timelike singularity requires a specification of boundary conditions that leads to a breakdown in
predictability.  In asymptotically flat space, a dynamical instability of the inner horizon of rotating and charged black holes is triggered
by generic small perturbations, cutting off the region behind it and supporting the  strong cosmic censorship conjecture \cite{flat space
blueshifts,numerical instability}.   Numerical and analytic studies of shells of matter in BTZ black holes suggest  an instability at the
inner horizon \cite{mass inflation papers}.   We will show that a generic scalar perturbation has divergent stress-energy on the inner
horizon which signals that the inner horizon is unstable to  scalar perturbations.

\noindent {\bf $\bullet$~Excursions in the complex plane: } Since we expect something to break down at the Cauchy horizon, we will
principally be interested in coordinate patches for a  set of regions that do not cross it. We focus on the regions $1_{+ \pm}$ and $2_{+
\pm}$ where
\begin{eqnarray}
T_1+X_1 &=& \sqrt{u} \exp \Bigl( \frac{r_+}{\Lambda} \phi_+ \Bigr) ,
\labell{embedbtz1} \\
T_2+X_2 &=& \sqrt{u-1} \exp \Bigl( \kappa_+ t - \frac{r_-}{\Lambda} \phi_+ \Bigr) , \labell{embedbtz2}
\end{eqnarray}
and $T_1 + X_1$ is always positive. We can now move from region to region by taking excursions in the complexified BTZ coordinate plane. In
region $1_{++}$ (the right asymptotic region), $u>1$ and $t$ is real.    To move to $2_{++}$ we first take $0<u<1$.   The square root in the
embedding equation (\ref{embedbtz2}) introduces a factor of $i$, so to keep the embedding real, and the sign of $T_2+X_2$ positive, we must
take $t\rightarrow t - i \beta_* /4$, where $\beta_* = 2 \pi / \kappa_+$ is the inverse Hawking temperature of the black hole. We have
selected a specific branch for continuing $t$, but there is nothing in the embedding equations that prevents us from taking a shift of $+3i
\beta_* /4$ or any other shift separated by a factor of $\beta_*$. Later on, we will see that dynamical issues will reduce this freedom to a
single choice of branch.  Since the shift in the complex direction is always a constant, and the $t$ dependence in the metric is always in
the form of a $dt$ term, it does not affect the metric.

As is well known, BTZ coordinates (\ref{untwisted btzmetric}) nominally break down at the outer horizon of the black hole and require some
sort of continuation to cover the spacetime.    If we are only interested in physics within one of the regions $A_{\pm\pm}$ we could define
separate BTZ-like coordinate patches in each region, defined by equations like (\ref{embedbtz1}, \ref{embedbtz2}).   This amounts to naively
setting $r < r_+$ in (\ref{untwisted btzmetric}).  However, if we start in one region and attempt to move into another while maintaining the
embedding equations (\ref{embedbtz1}, \ref{embedbtz2}), the different regions appear as different sections of a {\it complexified} BTZ
coordinate plane each having a real metric.  This perspective has been used in BTZ and other backgrounds \cite{Hemming:2002kd, kos,
leviross, Balasubramanian:2003kq, Fidkowski:2003nf}. Examining the relation between BTZ and Kruskal coordinates for the black hole provides
further insight into how this works.   It is not possible to find a coordinate system that covers both the outer and inner horizon. One set
of Kruskal coordinates, which extends from the AdS boundary up to, but not including the Cauchy horizon, is related to the BTZ coordinates
(\ref{untwisted btzmetric}) by \cite{bthz}
\begin{eqnarray}
2_{++}:~~~r_-<r\leq r_+\left\{ \begin{array}{ll}  X_+ = \biggl[ \biggl( \frac{r_+-r}{r+r_+} \biggr) \biggl( \frac{r+r_-}{r-r_-}
\biggr)^{r_-/r_+}
 \biggr]^{1/2} \sinh \kappa_+ t \\
T_+ = \biggl[ \biggl( \frac{r_+-r}{r+r_+} \biggr) \biggl( \frac{r+r_-}{r-r_-} \biggr)^{r_-/r_+} \biggr]^{1/2} \cosh \kappa_+ t
\end{array} \right. \labell{btzcoords1} \\
1_{++}:~~~r_+\leq r< \infty \left\{ \begin{array}{ll}  X_+ = \biggl[ \biggl( \frac{r-r_+}{r+r_+} \biggr) \biggl( \frac{r+r_-}{r-r_-}
\biggr)^{r_-/r_+}
 \biggr]^{1/2} \cosh \kappa_+ t \\
T_+ = \biggl[ \biggl( \frac{r-r_+}{r+r_+} \biggr) \biggl( \frac{r+r_-}{r-r_-} \biggr)^{r_-/r_+} \biggr]^{1/2} \sinh \kappa_+ t
\end{array} \right. \labell{btzcoords2}
\end{eqnarray}
The future and past event horizons occur at  $X_+ = \pm T_+$.    This would usually be described as a choice of different BTZ coordinate
patches for each region.   Here, we instead interpret passing through the horizon as a shift in $t$ by $-i\beta_*/4$, which exchanges the
$\sinh$ and $\cosh$ in (\ref{btzcoords1}) and (\ref{btzcoords2}).

In a similar way, to reach $1_{+-}$ from $1_{++}$ we take $t\rightarrow t - i \beta_* /2$, and to reach $2_{+-}$ we take $t\rightarrow t-3 i
\beta_* /4$. From (\ref{embedbtz2}) we can describe $1_{++}$ using either $t$ or $t+i n \beta_*$, where $n$ is an integer. A similar formula
can be found for the other regions. Later on, we will see that this freedom will be reduced to a single  choice of branch.

The Euclidean continuation of this spacetime is achieved by taking $t=-i\tau$ and continuing  $r_- = -i \tilde{r}_-$  to keep the metric
real:
\begin{equation}
ds_E ^2 = \frac{(r^2-r_+^2)(r^2+\tilde{r}_-^2)}{\Lambda ^2 r^2} d\tau^2 + \frac{\Lambda ^2 r^2}{(r^2-r_+^2)(r^2+\tilde{r}_-^2)}dr^2+r^2
\Bigl( d\phi_+ - \frac{\tilde{r}_-}{\Lambda r_+ r^2} (r^2-r_+^2) d\tau \Bigr)^2 . \label{twisted euclidean metric}
\end{equation}
The range of $r$ is now $r_+ < r < \infty$ and to avoid a conical singularity at $r=r_+$, $\tau$ must be periodic with period $\beta_*$.
Intriguingly, a discrete motion by $\tau=-\beta_* /2$ in the complexified BTZ time in effect takes us between disconnected components of the
Lorentzian spacetime boundary.

\subsection{Coordinate systems}
Throughout the paper we will switch between various appropriate coordinate systems. For reference we define them here and where appropriate
give their values in the different regions of interest. We have already defined the BTZ coordinates.  Tortoise coordinates in region
$1_{++}$ are given by
\begin{equation}
r_*=\frac{1}{2 \kappa_+} \ln(u-1)-\frac{1}{2 \kappa_-}\ln u ~~~~;~~~~ U=t-r_* ~~~~;~~~~ V=t+r_* \, . \labell{tort coords}
\end{equation}
In region $1_{++}$, the location of the past (future) event horizon in these coordinates is $V=- \infty$ ($U=\infty$).    In region
$2_{++}$, $r_*$ acquires an imaginary part by definition, and $t$ shifts into the complex plane as described above.   It will be useful
later to isolate the real parts of these coordinates so we define
\begin{equation}
\tilde{r}_* = \frac{1}{2 \kappa_+} \ln(1-u)-\frac{1}{2 \kappa_-}\ln u ~~~~;~~~~
 \tilde{U} = t - \tilde{r}_*
 ~~~~;~~~~
 \tilde{V} = t + \tilde{r}_*
 \labell{innertortcoords}
\end{equation}
Here $\tilde{r}$ is the real part of $r_*$ and $t$ does not include the finite shift of $-i\beta_* /4$ in region $2_{++}$. In these
coordinates, the left (right) Cauchy horizon is $\tilde{U}=- \infty \, (\tilde{V} = \infty)$.

As described in \cite{bthz} there are two sets of Kruskal coordinates, one covering the outer horizon but not the inner horizon, and vice
versa.   The outer coordinates are  given by (\ref{btzcoords1}, \ref{btzcoords2}). From these, we can construct null Kruskal coordinates
\begin{equation}
U_+=X_+ -T_+ ~~~~;~~~~V_+=X_++T_+ \,. \labell{nullouterkruskal}
\end{equation}
 These coordinates break down at the inner horizon as they diverge
and the metric has vanishing determinant. In terms of these coordinates we have
\begin{eqnarray}
&  1_{++}: ~V_+,U_+ >0 \ \ \ \ \  & 1_{+-}:~V_+,U_+<0 \nonumber \\
&  2_{++}: ~U_+<0, \ V_+>0 \ \ \ \ \  &  2_{+-}:~V_+<0, \ U_+>0
\end{eqnarray}
and the past and future event horizons occur at $V_+=0$ and $U_+=0$ respectively.

Kruskal coordinates that cover the inner horizon (but not the outer horizon) can also be defined and their relation with BTZ coordinates in
the region $2_{++}$, $r_-\leq r < r_+$ are given by
\begin{eqnarray}
X_-= - \biggl[ \biggl( \frac{r-r_-}{r+r_-} \biggr) \biggl( \frac{r_++r}{r_+-r} \biggr)^{r_+/r_-} \biggr]^{1/2} \sinh(\kappa_- t)
\\
T_-=  \biggl[ \biggl( \frac{r-r_-}{r+r_-} \biggr) \biggl( \frac{r_++r}{r_+-r} \biggr)^{r_+/r_-} \biggr]^{1/2} \cosh(\kappa_- t),
\end{eqnarray}
\begin{equation}
\phi_- = \bar{\phi} - \Omega_C t , \labell{phiminus}
\end{equation}
where
\begin{equation}
\Omega_C = {r_+ \over r_-\Lambda} \labell{innerangvel}
\end{equation}
 is the angular velocity of the inner horizon. The angular coordinate $\phi$ has be redefined to obtain
well-defined metric components at the inner horizon. From these, we can construct null Kruskal coordinates
\begin{equation}
U_-=X_- -T_- ~~~~;~~~~V_-=X_-+T_- \,. \labell{nullinnerkruskal}
\end{equation}
The left (right) Cauchy horizon in region $2_{++}$ is the line $U_-=X_- - T_-=0$ ($V_-=X_- + T_-=0$).

\section{The event horizon} \label{sec-eh}

The first goal of this paper is to demonstrate that the Cauchy horizon of the rotating BTZ black hole is unstable to small scalar
perturbations.     To show this we construct classical scalar wavepackets,  the stress tensor invariants of which are finite at the outer
horizon, but nevertheless diverge due to focusing effects at the inner horizon.  In this section we show that that standard basis of plane
wave normalizable modes in the asymptotic region $1_{++}$ is singular at the event horizon.   The divergence arises from an interaction
between the ingoing and outgoing pieces of the mode.  We then show how to construct wavepackets with regular behaviour at the outer horizon.
These wavepackets can then be used to diagnose potential blueshifting instabilities at the inner horizon.

\subsection{The need for wavepackets} \label{subsec-wp}
The plane-wave normalizable modes in region $1_{++}$ are  \cite{Keski-Vakkuri:1998nw}
\begin{equation} \label{eskos modes}
\phi_n(x) = e^{-i\omega t+ i n \phi_+ + i \Omega_H n t} (u-1)^\alpha u^{-h_+-\alpha} F(\alpha+\beta+h_+,\alpha-\beta+h_+;2h_+;u^{-1}) ,
\end{equation}
where $F$ is a hypergeometric function,
\begin{eqnarray}
\alpha &=& \frac{i \Lambda^2}{2 (r_+^2-r_-^2)} (r_+ \omega - r_- n/\Lambda) \nonumber \\
\beta &=& \frac{i \Lambda^2}{2 (r_+^2-r_-^2)} (r_- \omega -
r_+ n/\Lambda)  , \nonumber \\
h_{\pm} &=& \frac{1}{2} (1 \pm \nu) = \frac{1}{2} (1 \pm \sqrt{1+ m^2}) ,
\end{eqnarray}
and $m$ is the mass of the scalar field. This representation of the hypergeometric function is well adapted to describe the behavior on the
boundary, and one can readily verify that $\phi_n$ decays as $r^{-2h_+}$ there. To examine the behavior near the event horizon we use the
linear transformations of the hypergeometric function to obtain \cite{as}
\begin{eqnarray}
\phi_n(x) = e^{-i\omega t+ i n \phi_+ + i \Omega_H n t} [ A (u-1)^\alpha u ^\beta F(\alpha+\beta+h_+,\alpha+\beta+h_-;2\alpha+1;1-u) + \nonumber \\
B(u-1)^{-\alpha} u ^{-\beta} F(-\alpha-\beta+h_+,-\alpha-\beta+h_-;-2\alpha+1;1-u) ] , \labell{plane mode}
\end{eqnarray}
where
\begin{eqnarray}
A= \frac{ \Gamma(2h_+) \Gamma(-2\alpha)}{\Gamma(-\alpha-\beta+h_+) \Gamma(-\alpha+\beta+h_+)} ~~~;~~~ B=\frac{ \Gamma(2h_+)
\Gamma(2\alpha)}{\Gamma(\alpha+\beta+h_+) \Gamma(\alpha-\beta+h_+)} . \labell{AB}
\end{eqnarray}
The coefficient $A$ multiplies the outgoing part of the mode while $B$ multiplies the ingoing part.  Any normalizable mode solution of the
wave equation in the BTZ background will have both outgoing and ingoing pieces because of backscattering from the geometry.   By contrast in
asymptotically flat space we could have chosen a purely ingoing normalizable solution.

Near the event horizon ($u=1$), the leading terms in the solution are
\begin{equation} \label{plane mode 2}
\phi_n(x)\approx e^{-i\omega t+ i n \phi_+ + i \Omega_H n t} [ A (u-1)^\alpha +B(u-1)^{-\alpha} ].
\end{equation}
An invariant for checking the behavior of the field at any point is the trace of its stress-energy tensor.  The stress-energy tensor for a
minimally coupled scalar field is \cite{bd}
\begin{equation}
T_{\mu \nu} = \frac{1}{2} \Bigl( \partial_\mu \phi \partial_\nu \phi ^* + \partial_\nu \phi \partial_\mu \phi ^* \Bigr) - \frac{1}{2} g_{\mu
\nu} \Bigl( g^{\rho \sigma} \partial_\rho \phi \partial_\sigma \phi ^* + m^2 |\phi| ^2 \Bigr) .
\end{equation}
Taking the trace we obtain (for three spacetime dimensions)
\begin{equation}
T_\mu ^\mu = -\frac{1}{2} g^{\mu \nu} \partial_\mu \phi \partial_\nu \phi ^* -\frac{3}{2} m^2 |\phi|^2 .
\end{equation}
$\phi_n$ is well behaved on the event horizon, so any divergent behavior will come from the first part of the trace (note that this is also
of the same form as the kinetic term in the Lagrangian). Using our mode solution (\ref{plane mode 2}) for real $\omega$ we find
\begin{eqnarray}
g^{\mu \nu} \partial_\mu \phi_n \partial_\nu \phi_n ^* & \approx &    -2 [AB^* (u-1) ^{2\alpha} +BA^* (u-1)^{-2\alpha}]
{ \Lambda^2 \over (r_+^2-r_-^2) (r^2 - r_+^2)}(r_+ \omega - r_- n/\Lambda)^2  \nonumber \\
& + &\frac{n^2}{r_+^2-r_-^2}|\phi_n|^2 . \labell{stress1}
\end{eqnarray}
Since this expression is independent of $t$, the stress tensor diverges everywhere along the event horizon, which is reached by by taking $r
\to r_+$ while $t \to \pm \infty$.

Notice that the divergent terms depend on a product of $A$ and $B$ and hence involve an interaction between the outgoing and ingoing parts
of the solution.  This is the classic phenomenon of large center of mass energies developing in in-out collisions close to a black hole
horizon.  In asymptotically flat space, it is possible to choose purely ingoing boundary conditions at the event horizon but the the AdS
geometry does not allow this for normalizable solutions.   Purely ingoing modes will either be non-normalizable in the sense of diverging at
the AdS boundary or will involve the complex frequencies.    The latter solutions, the quasi-normal modes, are determined by setting the
outgoing coefficient $A$ in (\ref{AB}) to zero.  (See  \cite{quasinormal mode papers} for other studies of quasinormal modes in AdS black
holes.)  This is achieved by choosing the complex frequencies
\begin{eqnarray} \label{qnm}
\omega_L = -2i (h_+ +k)(r_+-r_-)/\Lambda^2 + n/\Lambda , \\
\omega_R= -2i(h_++k)(r_++r_-)/\Lambda^2-n/\Lambda ,
\end{eqnarray}
where $k=0,1\ldots$ and $n$ is the angular frequency in the mode (\ref{plane mode}).   Mode solutions with these frequencies decay
exponentially in the future, reflecting the infall into the horizon, but they grow exponentially in the past.  Due to this behavior  we find
a finite trace for the stress tensor on the future horizon but a divergence on the past horizon.  Therefore, the quasinormal modes do not
form a suitable set of perturbations that are regular in the black hole exterior.   Below we will construct wavepackets with the necessary
regularity and show that they are dominated at late times by quasinormal behavior.

\subsection{The invariants} \label{subsec-ehinv}

We have not yet fixed any normalization on the modes so we are free to multiply by an overall constant. For ease of computations later we
divide the solutions by an overall factor of $A$, giving
\begin{equation}
\phi_n(x)\approx e^{-i\omega t+ i n \phi_+ + i \Omega_H n t} [  (u-1)^\alpha +\frac{B(\omega)}{A(\omega)}(u-1)^{-\alpha} ] ,
\end{equation}
for the near horizon behavior. One can view the ratio $B/A$ as the dispersion for the ingoing wave as a result of scattering from the
gravitational well of AdS. From (\ref{tort coords}), near the outer horizon $r_* \approx (1/2\kappa_+) \ln(u-1)$. The mode solution near the
horizon can be approximated as
\begin{equation} \label{near horizon mode}
\phi_n(x) \approx e^{in \phi_+} \Bigl( e^{-i \omega U} e^{i \Omega_H n U} + \frac{B(\omega)}{A(\omega)} e^{-i\omega V} e^{i\Omega_H n V}
\Bigr),
\end{equation}
where $\Omega_H=\frac{r_-}{r_+ \Lambda}$ is the angular velocity of the outer horizon and we are using the tortoise coordinates defined in
(\ref{tort coords}) . To form a wavepacket we sum over modes of different $\omega$ with some weighting kernel $H(w)$. Define the functions
\begin{equation}
f(U) = \int _{-\infty} ^\infty d \omega ~ e^{-i \omega U} H(\omega) , \labell{f}
\end{equation}
\begin{equation}
g(V) =\int _{-\infty} ^\infty d \omega ~ e^{-i \omega V} \frac{B(\omega)}{A(\omega)}H(\omega) .
 \labell{g}
\end{equation}
Near the event horizon, the wavepacket is
\begin{equation}
\phi_{wp} (x) \approx e^{in\phi_+} [ f(U) e^{i\Omega_H n U} + g(V) e^{i\Omega_H n V} ]. \labell{near horizon wavepacket}
\end{equation}
Here $f(U)$ is the outgoing part of the wavepacket and $g(V)$ is the ingoing part.   We need $f$ to be small at the future horizon and $g$
to be small at the past horizon in order to have a finite stress-energy.
 We wish to compute the trace of the stress-energy tensor with our wavepacket. As before, any
divergences come from the kinetic term.   Setting $f' = df/dU$ and $g' = dg/dV$, we obtain
\begin{eqnarray}
g^{\mu \nu} \partial_\mu \phi_{wp} \partial_\nu \phi_{wp} ^* \approx \frac{2 \Lambda^2 r_+ ^2}{(r_+^2-r_-^2)^2} e^{\kappa_+ (U-V)} & \biggl(
& e^{i\Omega_H n (U-V)} [ -f' g'^* + i \Omega_H n f' g^* - i \Omega_H n f g'^* - \Omega_H^2n^2fg^* ]
\nonumber  \\
& + & {\rm c.c.} \biggr) +\frac{n^2}{r_+^2 - r_-^2} |\phi_{wp}|^2. \labell{wp eh inv}
\end{eqnarray}
Despite the apparent divergence as $e^{\kappa_+U}$ ($e^{-\kappa_+ V}$) as we approach the future (past) event horizon, this stress tensor
trace can be in fact be finite if the kernel $H(\omega)$ in (\ref{f}, \ref{g}) is appropriately chosen.  We turn to this below.

\subsection{Computation of the wavepackets and their behavior at the event horizon}
\label{subsec-ehwp}

To get a finite stress tensor at the event horizon the kernel $H(\omega)$ (\ref{f}, \ref{g}) must be chosen to suppress the
$e^{\kappa_+(U-V)}$ that appears as an overall factor in (\ref{wp eh inv}).   We will not try to find a general basis of wavepackets
satisfying this condition, but will instead give sufficient conditions on $H(\omega)$ that lead to regular wavepackets.

Let us assume that $H(\omega)$ is a meromorphic function of $\omega$ which falls off sufficiently quickly at infinity to allow use of
contour integration in (\ref{f}, \ref{g}).  The potential divergence at the past event horizon ($V \to -\infty$) in (\ref{wp eh inv}) occurs
because of the overall factor of $e^{-\kappa_+ V}$.    To suppress this, $g(V)$ must decay sufficiently quickly as $V \to -\infty$.
(Similarly $f(U)$ must decay sufficiently quickly as $U \to \infty$ for the stress tensor to be finite on the future horizon.)   To consider
the behavior of $g(V)$ on the past horizon, we can compute the integral (\ref{g}) by closing the contour in the upper half plane (UHP) for
convergence on the semicircle at infinity.  Poles can arise from the the factor $B/A$ and from the kernel $H$.  From (\ref{AB}) we have
\begin{equation}
\frac{B(\omega)}{A(\omega)} = \frac{ \Gamma(2 \alpha) \,  \Gamma (h_+ - \alpha -\beta) \,  \Gamma (h_+ - \alpha + \beta)}{\Gamma(-2 \alpha)
\,  \Gamma (h_+ + \alpha +\beta) \, \Gamma (h_+ + \alpha - \beta ) }. \labell{B/A}
\end{equation}
The term $\Gamma(2\alpha)$ in the numerator gives rise to poles at
\begin{equation}
\omega = i k \kappa_+ + \frac{r_-}{r_+} n , \labell{B/A poles}
\end{equation}
where $k=1,2,\ldots$ (the $k=0$ pole does not contribute because the $\Gamma(-2\alpha)$ in the denominator cancels it).   These poles are in
the upper half plane and contribute.   (The other Gamma functions in the numerator give rise to poles in the lower half plane (LHP) and do
not contribute.)   The dominant contribution will be from the $k=1$ pole as $V\rightarrow -\infty$ which leads to a behavior of $g(V) \sim
e^{\kappa_+ V}$.   This precisely cancels the overall divergent factor $e^{-\kappa_+ V}$ in (\ref{wp eh inv}) that appears at the past
horizon $V \to -\infty$.    Therefore, to have a finite stress tensor trace on the past horizon it suffices to require that any poles in $H$
that lie in the UHP have $Im (\omega) \geq \kappa_+$.   A similar analysis on the future horizon for $f(U)$ leads to the condition that any
poles in $H$ that lie in the LHP have $Im (\omega) \leq -\kappa_+$.  In summary, for the stress tensor to have a finite trace it is
sufficient that poles in $H(\omega)$ satisfy the condition
\begin{equation}
|Im(\omega)| \geq \kappa_+ \, . \labell{condition1}
\end{equation}
(For simplicity we have assumed that $H(\omega)$ does not cancel any poles in $B/A$, and that there are no double poles. These situations
can be handled in a straightforward manner.)  Other invariants constructed out of the stress tensor are also expected to be finite under
these conditions.

We can also show that on the future horizon ($U \to \infty$) the late time behavior ($V \to \infty$) of the regular wavepackets is dominated
by quasinormal modes.   The outgoing part of the wavepacket, which is controlled by $f(U)$, is exponentially suppressed by construction; so
we are interested in the behavior of the ingoing part, $g(V)$.   To compute the late time behavior of this part, i.e., as $V \to \infty$ in
(\ref{g}), we close the contour in the LHP.  Then, in addition to any poles in $H(\omega)$ we pick up the poles from the latter two gamma
functions in the numerator of  (\ref{B/A}). The poles in $\Gamma(h_+ - \alpha - \beta)$ occur at
\begin{equation}
\omega = \omega_L = -2i (h_+ +k)(r_+-r_-)/\Lambda^2 + n/\Lambda ~~~;~~~ k = 0,1,\ldots \, ,
\end{equation}
The poles in $\Gamma(h_+ - \alpha +\beta)$ occur at
\begin{equation}
\omega = \omega_R= -2i(h_++k)(r_++r_-)/\Lambda^2-n/\Lambda
 ~~~;~~~ k = 0,1,\ldots
 \,  .
\end{equation}
>From the residues at these poles we see that at late times on the future horizon, the wavepacket is dominated by a sum of quasinormal modes
as we should expect.

\section{Continuing to the other regions and excursions in the complex plane}
\label{sec-cont}

In the previous section we found conditions on wavepackets in the $1_{++}$ asymptotic region which led to a well-defined stress tensor on
the outer horizon.  In order to study whether such a packet can seed an instability of the inner horizon we must continue the solution to
other regions of the black hole.    Requiring single valuedness, continuity and positive energy places conditions on both the kernel that
determines the wavepacket and the locations of branch cuts in the complex Kruskal coordinate plane.  Below we will determine these
conditions and translate them back to BTZ coordinates.

\subsection{Continuity of the modes and the wavepacket}

First we examine continuity of the packets across the event horizon.  Near the event horizon in the region $1_{++}$ the relation between BTZ
and outer Kruskal coordinates (\ref{nullouterkruskal}) is
\begin{equation}
t=\frac{1}{2 \kappa_+} ( \ln V_+ - \ln U_+) ~~~~;~~~~ u-1 \approx \frac{4r_+^2}{r_+^2-r_-^2} \biggl( \frac{r_+ -r_-}{r_+ + r_-}
\biggr)^{r_-/r_+} U_+V_+ \, . \labell{near event horizon kruskal 2}
\end{equation}
Using these transformations the null tortoise coordinates (\ref{tort coords}) are
\begin{equation}
U= -\frac{1}{\kappa_+} [ \ln \gamma + \ln U_+]   ~~~~;~~~~ V= \frac{1}{\kappa_+} [ \ln \gamma + \ln V_+] ,  \labell {kruskal v transform}
\end{equation}
where
\begin{equation}
\gamma ^2 = \frac{4r_+^2}{r_+^2-r_-^2} \biggl( \frac{r_+ -r_-}{r_+ + r_-} \biggr)^{r_-/r_+} \, . \labell{gammadef}
\end{equation}
We can find the wavepacket in the other regions all the way up to the inner horizon by letting the Kruskal coordinates range over all real
values from $-\infty$ to $\infty$.  First,  consider the modes (\ref{near horizon mode}) as we pass into either region 2 from the asymptotic
region $1_{++}$.    As we pass through the future horizon at $U_+ = 0$, the outgoing ($U$ dependent) part of the solution is
\begin{equation}
e^{-i\omega U+i \Omega_H n U} \sim e^{-i/\kappa_+ \omega \ln U_+ + i/\kappa_+ \Omega_H n \ln U_+} \, .
\end{equation}
This is discontinuous across $U_+=0$ because of the branch cut in the log in the exponent.  Similarly, the ingoing ($V$ dependent) part of
the mode is discontinuous across the the past horizon ($V_+ = 0$).    We have already shown that the stress tensor for the mode solutions
diverges on the event horizon precisely because the outgoing (ingoing) part is non-vanishing on the future (past) horizon.   The latter
problem was cured by making wavepackets in which the outgoing (ingoing) part vanished on the future (past) horizon.   We will see below that
continuity of the wavepackets is linked to the finiteness and continuity of the stress tensor.

Near the event horizon the wavepacket is given by (\ref{near horizon wavepacket}).  We examine the outgoing ($f(U)$) and ingoing  ($g(V)$)
parts separately.   Near the future event horizon ($U_+=0$) the ingoing part of the wavepacket is manifestly continuous and differentiable.
The dangerous part is the outgoing piece $f(U)$.  The contour integral defining $f(U)$ (\ref{f}) receives contributions from poles in the
kernel $H(\omega)$ that lie in the LHP.  We parametrize the location of these poles as $\omega = -a i \kappa_+ \mp b \kappa_+$ where $a,b$
are positive, real constants. Then, near the future horizon, the outgoing part of the wavepacket behaves as
\begin{equation}
e^{i\Omega_H n U} f(U) \sim e^{i(\Omega_H n \pm b \kappa_+) U -a \kappa_+ U} \sim \exp \biggl[ -i \biggl( \frac{\Omega_H n}{\kappa_+} \pm b
\biggr)(\ln \gamma +\ln U_+) + a(\ln U_+ + \ln \gamma) \biggr] .
\end{equation}
When $U_+<0$ the right hand side of this equation becomes
\begin{equation}
 \exp \biggl[ \biggl( \frac{\Omega_H n}{\kappa_+} \pm b \biggr) m\pi + iam\pi \biggr] \exp \biggl[ -i \biggl(
\frac{\Omega_H n}{\kappa_+} \pm b \biggr)(\ln \gamma +\ln (-U_+)) + a(\ln (-U_+) + \ln \gamma) \biggr] ,
\end{equation}
where $m$ represents the choice of branch in the log, which we will come back to shortly. The second exponential is the original wavepacket
again. Therefore, if $a>0$, the expression vanishes at $U_+=0$ and the wavepacket is continuous.   If $a\geq 1$ then the packet is
differentiable, and so on.   Similar issues of continuity and differentiability of wavepackets arise in studies of standard Schwarzschild
black holes \cite{unruh1976, Boulware:dm}.

Similarly, at the past horizon ($V_+=0$) the outgoing part of the wavepacket is manifestly continuous and differentiable, but the ingoing
piece ($g(V)$) requires analysis.   The contour integral defining $g(V)$ (\ref{g}) near the past horizon receives  contributions from the
poles in both  $B(\omega)/A(\omega)$ and the kernel $H(\omega)$ that lie in the upper half plane. The poles from $B/A$ (\ref{B/A poles})
lead to
\begin{equation}
e^{i\Omega_H n V} g(V) \sim \sum_{k=1} ^{\infty} Res_{\omega_k} (HB/A) e^{k \kappa_+ V} \sim \sum_{k=1} ^{\infty} Res_{\omega_k} (HB/A)
\gamma^k V_+^k ,
\end{equation}
as $V_+\rightarrow 0$. Since $k$ are positive integers these terms are $C^{\infty}$ at $V_+=0$.  Then, following the  analysis above, one
finds similar requirements for any poles of $H(\omega)$ in the UHP and a branch cut from $\ln V_+$.

One can  check that the trace of the stress tensor is continuous so long as $H(\omega)$ only has poles at $\omega_p$ with  $|Im(\omega_p)| >
\kappa_+$.  The requirement that the stress tensor is continuous is thus sufficient to ensure that the stress tensor is  finite and that the
wavepacket is continuous at the event horizon.

\subsection{The branch cuts and a positive energy condition}
As discussed above, the regular wavepackets have branch cuts in both the coordinates $U_+$ and $V_+$ that start  at the origin and extend
out to infinity.   We would like a physical criterion for placing the cuts in the complex plane.  Related issues arise in studies of
Schwarzschild black holes in asymptotically flat space and even in the Rindler description of Minkowski space.  In these cases, a method for
determining the placement of the branch cuts was given by Unruh \cite{unruh1976}.  Here we extend this procedure to the BTZ spacetime.
Essentially, the method involves using a condition of positive energy to select whether the cuts extend from the origin into the upper or
lower half Kruskal coordinate planes.    This is a discrete choice that does not depend on the angular momentum in the solution.   For ease
of computation, therefore, we will determine the location of the cuts in the non-rotating case with $r_- = 0$.    The result will apply also
to the rotating case.

Unruh's method relies on finding null vectors which become of the Killing type on the past and future event horizons. In Kruskal
coordinates, the non-rotating metric can be written \cite{bthz}
\begin{equation}
ds^2 = \Omega^2(r) \, dU_+ \, dV_+ + r^2 \, d\phi_+ ^2 ,
\end{equation}
\begin{equation}
\Omega^2(r) = \frac{(r+r_+)^2}{\kappa_+ ^2 \Lambda^2} ,
\end{equation}
and $r$, which is the BTZ radial coordinate, is implicitly defined in terms of $U_+$ and $V_+$ in each BTZ patch.  In the non-rotating case,
on the future horizon ($U_+=0$), $\partial_{V_+}$ becomes a Killing vector, and on the past horizon ($V_+=0$) $\partial_{U_+}$ becomes a
Killing vector. Following Unruh, on the past horizon we require
\begin{equation}
{\cal L}_{\partial_{U_+}} \tilde{\phi} = -i \tilde{\omega} \tilde{\phi}  ,
\end{equation}
for a wavefunction to have positive energy in the Kruskal sense. Here, ${\cal L}_{\partial_{U_+}}$ is the Lie derivative, $\tilde{\phi}$ is
a Kruskal mode, and $\tilde{\omega}>0$. Hence, on the past horizon we have $\tilde{\phi} \sim e^{-i \tilde{\omega} U_+}$. This wavefunction
is analytic in the lower half $U_+$ plane, and so we can characterize any positive energy mode by this analyticity. Going back to our modes
and wavepacket, this condition tells us to place the branch cut in the upper half $U_+$ plane. Carrying out a similar analysis on the future
event horizon, tells us that the branch cut in $V_+$ must also be in the UHP. With these branch choices, a BTZ mode with {\it any} $\omega$
is a positive energy Kruskal mode. We still have a choice as to the overall Riemann sheet that we are on, i.e.\ what the phase of $1$ is.
This choice, along with the choice of kernel, $H(\omega)$, determine whether the wavepacket is concentrated in one asymptotic region versus
the other, or in both.   For convenience we will choose the phase $1=e^{i 0}$, so that $\ln(1) = 0$.   (Given a kernel $H(\omega)$ different
sheets in which $\ln(1) = 2\pi l$ will result in packets localized in different ways.)

These branch cuts put a restriction on the BTZ coordinate time $t$ as well. From the coordinate transformations (\ref{btzcoords2}) in region
$1_{++}$  we have
\begin{equation}
U_+ = e^{-\kappa_+ t} P(r) ~~~~;~~~~ V_+ = e^{\kappa_+ t} P(r) \, , \labell{btzconstraint}
\end{equation}
where $P$ is a function of the BTZ radial coordinate.   Recall that in Sec.~\ref{sec-btz} we showed that one can move between different
regions of the BTZ geometry by making discrete translations in the complexified $t$ plane.  In particular, as far the as embedding equations
(\ref{embedding}) go translating $t$ by integral multiples of $2\pi i /\kappa_+ = i \beta_*$ brings us back to the same spacetime point.
However, notice that the cuts in $U_+$ and $V_+$ arising from the wavepacket, coupled with the relations (\ref{btzconstraint}), restrict the
imaginary parts of $t$.   Let $Im(t_\pm)$ denote the maximum and minimum values of the imaginary part of $t$ respectively.  Then to stay on
a single sheet of the log (namely to avoid the cuts in $U_+$ and $V_+$), we must have
\begin{equation} \label{t restriction}
Im(t_+) - Im(t_-) < 2 \pi / \kappa_+ = \beta_*
\end{equation}
Note that the inequality is a $<$, not a $\leq$.

\noindent {\bf $\bullet$~Relation to excursions in the complex plane: } We will now work out in detail  how these branch choices in the
Kruskal $U_+$ and $V_+$ planes affect the complexified BTZ coordinates.   In the asymptotic region $1_{++}$, $U_+,V_+ > 0$, so no choice of
branch is necessary. Near the event horizon the modes making up the wavepacket behave as in (\ref{near horizon mode}). In $1_{++}$ the
outgoing ($U$ dependent) term can be written
\begin{equation}
e^{-i\omega U+i \Omega_H n U} \sim e^{-(i/\kappa_+) \,  \omega \ln U_+ + (i/\kappa_+)  \, \Omega_H n \ln U_+} ~~,
\end{equation}
where we have left out a constant factor involving $\gamma$ (\ref{gammadef}). Passing through $U_+=0$ into $2_{++}$ (where $U_+ < 0$) we can
explicitly represent the choice of branch by writing the right hand side of the above equation as
\begin{equation} \label{going to region 2}
e^{-(i/\kappa_+) \omega \ln U_+ + (i/\kappa_+) \Omega_H n \ln U_+} = e^{-\frac{\pi \omega}{\kappa_+} + \frac{\pi \Omega_H n}{\kappa_+}}
e^{-(i/\kappa_+) \,  \omega \ln (-U_+) + (i/\kappa_+) \,  \Omega_H n \ln (-U_+)} ~~.
\end{equation}
In BTZ coordinates the factors on the right hand side coming from the branches of the log will get represented via both cuts in the
coordinate planes and by an imaginary shift in $t$ (as expected from the excursions in the complex plane that move between BTZ regions).

 To characterize this,
let us  write $t\rightarrow t - ic$ where $c$ is a real constant we wish to find. Rewriting (\ref{going to region 2}) in terms of BTZ
coordinates we find (note that $0<u<1$ since we are in region $2_{++}$ behind the horizon)
\begin{equation}
e^{-\frac{\pi \omega}{\kappa_+} + \frac{\pi \Omega_H n}{\kappa_+}} e^{-i/\kappa_+ \omega \ln (-U_+) + i/\kappa_+ \Omega_H n \ln (-U_+)} =
(1-u)^\alpha (-1)^\alpha e^{-i \omega (t-i c) + i \Omega_H n (t-ic)} ,
\end{equation}
where we have ignored terms involving the constant $\gamma$  (\ref{gammadef}) which do not affect our analysis. Similarly the ingoing ($V$
dependent) piece of the mode solution (\ref{near horizon mode}) leads to
\begin{equation}
e^{-i/\kappa_+ \omega \ln (V_+) + i/\kappa_+ \Omega_H n \ln (V_+)} = (1-u)^{-\alpha} (-1)^{-\alpha} e^{-i \omega (t-i c) + \Omega_H n
(t-ic)} ,
\end{equation}
with no phase shift on the left hand side because $V_+>0$ in both $1_{++}$ and $2_{++}$.   In order to match the Kruskal coordinate phase
shifts in BTZ coordinates we must have
\begin{equation}
(-1)^{\alpha} e^{-\omega c+\Omega_H n c} = e^{-\frac{\pi \omega}{\kappa_+} + \frac{\pi \Omega_H n}{\kappa_+}} ~~~~;~~~~ (-1)^{-\alpha}
e^{-\omega c+\Omega_H n c}=1 .
\end{equation}
Solving gives
\begin{equation}
(-1)^\alpha = e^{-\frac{\beta_*}{4} (\omega - \Omega_H n)}
  ~~~~;~~~~
 c= \beta_* /4 \, .
 \labell{-1}
\end{equation}
The first equation amounts to a choice of branch in $r_*$, the tortoise coordinate (\ref{tort coords}) constructed from BTZ coordinates. The
second equation, amounts to a statement that moving from the asymptotic region $1_{++}$ to the region $2_{++}$ behind the event horizon is
represented in BTZ coordinates as
\begin{equation}
t \to t - i \beta_*/4 ~~.
\end{equation}
In other words the ambiguity in shifting $t$ by integral multiples of $i\beta_*$ is fixed by the need to match the Kruskal branch cuts.
Similarly we find that moving from the right asymptotic region $1_{++}$ to the left asymptotic region $1_{+-}$ amounts to taking
$t\rightarrow t - i\beta_* /2$.

\section{The Cauchy horizon} \label{sec-ch}
Using the results above we propagate the mode solutions to the inner horizon and show that a  wavepacket that is regular at the outer
horizon nevertheless has divergent stress-energy on the inner horizon.   It is expected that this will lead to large backreaction and  an
instability in the spacetime.   The analysis of this section uses techniques similar to those in section \ref{sec-eh}.

To continue our mode solutions from the $1_{++}$ region to the $2_{++}$ region we must take account of the cuts and shifts in BTZ
coordinates that were described in the previous section.  Near the event horizon in the $2_{++}$ region the mode solutions are given by
\begin{eqnarray}
\phi_n(x) &= & e^{-i\omega (t-i\beta_*/4)+ i n \phi_+ +i\Omega_H n (t-i\beta_*/4)} \Bigl[  (u-1)^\alpha u ^\beta
F(\alpha+\beta+h_+,\alpha+\beta+h_-;2\alpha+1;1-u) + \nonumber \\
& & \frac{B}{A}(u-1)^{-\alpha} u ^{-\beta} F(-\alpha-\beta+h_+,-\alpha-\beta+h_-;-2\alpha+1;1-u) \Bigr] ~~,
\end{eqnarray}
where the factors in front come from the cuts and shifts described in the previous section.   In these coordiantes $u=1$ is the event
horizon while $u=0$ is the inner horizon.

To study the properties of the solution near the inner horizon we use the  linear transformation formulae for hypergeometric functions
\cite{as} to write them as functions of $u$.  Then the leading terms in $\phi_n$ near $u=0$ are
\begin{equation}
\phi_n^{2_{++}} \approx e^{in\phi_-} e^{-i \omega t +i \Omega_C n t} e^{-i \frac{\beta_*}{4} (in \Omega_H - i \omega) } \Bigl\{ [  D
(-1)^\alpha +  E (-1)^{-\alpha} ] \, u^{-\beta} + [  C (-1)^\alpha +  F (-1)^{-\alpha}] \, u^\beta \Bigr\} ,
\end{equation}
(We have switched to the angular coordinate $\phi_-$ (\ref{phiminus}) which is appropriate near the inner horizon.)  The coefficients, which
arise from the transformation formulae and $B/A$ (\ref{B/A}) are
\begin{eqnarray}
C&=&  \frac{\Gamma(2\alpha+1) \Gamma(-2\beta)}{\Gamma(\alpha -\beta +h_+) \Gamma(\alpha-\beta+h_-)}, \hspace{.5in} \\
D&=&  \frac{\Gamma(2\alpha+1) \Gamma(2\beta)}{\Gamma(\alpha +\beta +h_+) \Gamma(\alpha+\beta+h_-)}, \hspace{.5in} \\
E&=& \frac{ \Gamma(2\alpha)\Gamma(-\alpha -\beta +h_+) \Gamma(-2\alpha +1) \Gamma(2\beta)}{\Gamma(-2\alpha)
\Gamma(\alpha+\beta+h_+) \Gamma(\alpha-\beta+h_+) \Gamma(-\alpha +\beta +h_-)} , \\
F&=& \frac{ \Gamma(2\alpha)\Gamma(-\alpha +\beta +h_+) \Gamma(-2\alpha +1) \Gamma(-2\beta)}{\Gamma(-2\alpha) \Gamma(\alpha+\beta+h_+)
\Gamma(\alpha-\beta+h_+) \Gamma(-\alpha -\beta +h_-)} .
\end{eqnarray}
Near the inner horizon we can use the set of tortoise coordinates defined in (\ref{innertortcoords}). In terms of these coordinates,
 $\tilde{r}_* \approx -\frac{1}{2 \kappa_-} \ln u$ so $u^{\pm \beta} \approx \exp( \mp i \omega \tilde{r}_* \pm i
\Omega_C n \tilde{r}_* )$. Using our expression for $(-1)^\alpha$ (\ref{-1}) we find
\begin{equation} \label{region 2 mode}
\phi_n^{2_{++}} \approx e^{in\phi_-} \Bigl[ (D  e^{- \frac{\beta_*}{2} (\omega -n \Omega_H)} +E)e^{-i \omega \tilde{U} +i \Omega_C n
\tilde{U}} + (C e^{- \frac{\beta_*}{2} (\omega -n \Omega_H)}    + F )e^{-i \omega \tilde{V} +i \Omega_C n \tilde{V}} \Bigr] \, ,
\end{equation}
where $\tilde{U}$ ($\tilde{V}$) dependent terms are waves that move towards the right (left) Cauchy horizon in Fig.~\ref{fig:horizon}. Note
that even if the wavepacket had been purely ingoing at the outer horizon,  scattering from the geometry would have led to both left and
right moving pieces at the inner horizon.

Near the inner horizon the wavepacket is
\begin{equation} \label{inner horizon wp}
\phi_{wp} (x) \approx e^{in \phi_-} [ \tilde f (\tilde U) e^{i \Omega_C n \tilde U} + \tilde g (\tilde V) e^{i \Omega_C n \tilde V} ] ,
\end{equation}
where the rightmoving piece is
\begin{equation}
\tilde f (\tilde U) =  \int _{-\infty} ^\infty d\omega [ D(\omega) e^{- \frac{\beta_*}{2} (\omega -n \Omega_H)} + E(\omega) ] H(\omega)
e^{-i \omega \tilde U} , \labell{newf}
\end{equation}
and the leftmoving piece is
\begin{equation}
\tilde g (\tilde V) =  \int _{-\infty} ^\infty d\omega [ C(\omega)e^{- \frac{\beta_*}{2} (\omega -n \Omega_H)} + F(\omega) ] H(\omega) e^{-i
\omega \tilde V} . \labell{newg}
\end{equation}

Using the wavepacket we again find that any divergent contributions to the trace of the stress-energy tensor come from the kinetic term,
given by
\begin{eqnarray} \label{inner horizon invariant}
g^{\mu \nu} \partial_\mu \phi_{wp} \partial_\nu \phi_{wp} ^* \approx \frac{2 \Lambda^2 r_-^2}{(r_+^2 - r_-^2)^2} e^{\kappa_- (\tilde V -
\tilde U)} \Bigl( e^{i \Omega_C n (\tilde{U} - \tilde{V})} [ \tilde f' \tilde g'^* +i \Omega_C n \tilde f \tilde g'^* - i \Omega_C n \tilde
f' \tilde g ^* + \Omega_C ^2 n^2 \tilde f \tilde g^*] +c.c. \Bigr) \nonumber \\ - \frac{n^2}{r_+^2 -r_-^2} |\phi_{wp}|^2 , \hspace{3in}
\end{eqnarray}
where the primes are understood in the same way as they were for the event horizon computation. Thus, the overall factor causes a naive
divergence in the trace of the stress tensor  as $e^{\kappa_- \tilde{V}}$ ($e^{-\kappa_- \tilde{U}}$) as we approach the right (left) Cauchy
horizon.   As with the event horizon this potential divergence involves an interaction between right and left moving pieces.

Let us examine the behavior of the wavepacket on the left Cauchy horizon.  Assume as before that the kernel $H(\omega)$ decays sufficiently
quickly at large $\omega$ to use contour integration in (\ref{newf}) and (\ref{newg}).    At the left Cauchy horizon ($\tilde{U} \to -
\infty$), the potential divergence of the stress tensor can only be suppressed if the rightmoving piece falls off fast enough as $\tilde{U}
\to - \infty$.   To study this, we close the contour integral for $\tilde{f}(\tilde{U})$ in the UHP and get contributions from poles in
$D(\omega)$ arising from $\Gamma(2 \alpha+1)$ and $\Gamma(2\beta)$ factors. They occur at
\begin{eqnarray}
\omega_{1,k}&=& i\kappa_+ (k+1) + \frac{r_-}{r_+ \Lambda} n, \ \ \ k=0,1,\ldots ~~,\nonumber \\
\omega_{2,k}&=&  i \kappa_- k + \frac{r_+}{r_- \Lambda} n, \ \ \ k=0,1,\ldots  , \labell{dpoles}
\end{eqnarray}
respectively. Contributions to $\tilde{f}(\tilde{U})$ can also arise from the  poles in $E(\omega)$, but those that lie in the UHP are
identical to those in $D(\omega)$. In addition, continuity and finiteness of the stress tensor at the event horizon have imposed a condition
that any poles in $H(\omega)$ have an imaginary part with magnitude greater than $\kappa_+$. Therefore, the dominant behavior for the
rightmoving part of the wavepacket, $\tilde{f}(\tilde{U})$ on the left Cauchy horizon ($\tilde{U} \to - \infty$) (ignoring oscillating
parts), is given by the residues associated with the $k=0$ poles in (\ref{dpoles}).    We write this as
\begin{eqnarray} \label{past inner horizon behavior}
\tilde{f} (\tilde{U}) &\sim&  (Res(EH) + Res(DHe^{- \frac{\beta_*}{2} (\omega -n \Omega_H)})) |_{\omega=\omega_{2,0}}  \nonumber \\
&+& (Res(EH)+ Res(DHe^{- \frac{\beta_*}{2} (\omega -n \Omega_H)}))e^{\kappa_+ \tilde{U}} |_{\omega=\omega_{1,0}} ~~.
\end{eqnarray}
These terms do not decay fast enough on the left Cauchy horizon at $\tilde{U} \to - \infty$ to suppress the divergence in the trace of the
stress tensor coming from the overall factor in  (\ref{inner horizon invariant}).

Similarly the potential divergence in  (\ref{inner horizon invariant}) at the right Cauchy horizon ($\tilde{V} \to \infty$) can only be
suppressed by a sufficiently rapid decay of the leftmoving part $\tilde g(\tilde V)$ of the wavepacket.   Closing the contour integral
defining (\ref{newg}) in the LHP we obtain the leading contributions
\begin{eqnarray}
\tilde{g}(\tilde{V}) &\sim&  [Res(FH)+Res(CHe^{- \frac{\beta_*}{2} (\omega -n \Omega_H)}] |_{\omega_{2,0}} \nonumber \\
&+& \exp\Bigl[ -2h_+ \frac{ (r_+ +r_-)}{\Lambda^2}\tilde V \Bigr] Res(FH)  |_{\omega_{3,0}} ~~, \label{future inner horizon behavior}
\end{eqnarray}
where $\omega_{2,k}$ is given above in (\ref{dpoles}) and
\begin{equation}
\omega_{3,k} = \omega_R= -2i(h_++k)(r_++r_-)/\Lambda^2-n/\Lambda , \ \ k=0,1,\ldots \ \ .
\end{equation}
Interestingly the frequencies $\omega_R$ are precisely those of half the quasinormal modes.  While we only included the $\omega_{i,0}$
contributions in the above analysis of the leading terms in the trace of the stress tensor, the higher $k$ frequencies  contribute to
subleading behaviors.   The fact that the quasinormal frequencies contribute only in the leftmoving part of the wavepacket emphasizes again
that the perturbation we are constructing is not symmetric between the two asymptotic regions.  Thus, the effect we are seeing does not
arise from a collision of a symmetric packet, but rather due to the effects of scattering from the geometry and gravitational focusing. As a
check on this, one can easily show that even a (non-normalizable) purely ingoing wavepacket at the event horizon leads to divergent behavior
on both the left and right Cauchy horizons.

Indeed it is easy to numerically construct a wavepacket with a finite stress tensor on the event horizon, that is exponentially suppressed
in the other asymptotic region and nevertheless has a divergent stress tensor on the Cauchy horizon. It is commonly believed that such
divergent behavior will cause a gravitational instability at the inner horizon. This is expected to lead to the formation of a singularity
and the preservation of cosmic censorship.

\section{AdS/CFT correlators and detecting the instability} \label{sec-adscorr}
We have shown how to construct wavepackets with a stress tensor that is finite at the outer horizon, but diverges at the inner horizon.
Since the BTZ black hole is asymptotically AdS, there is a description of the spacetime in a dual CFT
\cite{Maldacena:1997re,Gubser:1998bc,Witten:1998qj}.   In this section we will explore whether and how the CFT is sensitive to the
instability at the inner horizon. (Previous work, from a variety of perspectives, on the holographic encoding of physics behind a horizon
includes \cite{simongary,louko,vijaysimon,marolf,maldacena,veronika}.)

In Lorentzian signature the AdS/CFT correspondence is formulated as \cite{vijay1,vijay2}
\begin{equation} \label{ads/cft}
\langle \phi_n | T \exp i \int_\partial \phi_0 {\cal O} | \phi_n \rangle_{CFT} = Z(\phi_0) ,
\end{equation}
where $Z$ is the string or supergravity partition function with the boundary condition that $\phi \rightarrow \phi_0$ on the boundary, $T$
is the time ordering symbol and ${\cal O}$ is the operator dual to $\phi$. The state $ | \phi_n \rangle $ is dual to any normalizable
component in the bulk field $\phi (x) $ \cite{vijay1, vijay2}. We can compute $n$-point functions of ${\cal O}$ by taking functional
derivatives with respect to $\phi_0$ and then setting $\phi_0=0$ in the usual way.

As is well known, in Lorentzian signature we have two possible types of modes: normalizable and non-normalizable. Non-normalizable modes are
dual to sources in the boundary theory and are also present in the Euclidean formulation of AdS/CFT. A non-normalizable mode can be
expressed as
\begin{equation}
\phi_{nn} (x) = \int d {\bf b} K(x,{\bf b}) \phi_0 (\bf{b}) ,
\end{equation}
where $K(x,\bf{b})$ is the bulk-boundary propagator, $x$ are coordinates representing the bulk spacetime and $\bf{b}$ lies in the boundary.
In the Euclidean formulation of AdS/CFT, these are the only modes present, and by plugging this expression into (\ref{ads/cft}) one can
build up amplitudes by stringing together bulk-boundary propagators in the appropriate way \cite{Witten:1998qj, Gubser:1998bc}.

Since our instability is a result of a wavepacket we must include normalizable modes and write
\begin{equation}
\phi(x) = \phi_{wp} (x) + \int d {\bf b} \, K(x,\bf{b}) \, \phi_0 (b)
\end{equation}
as the expression to be used for computing correlators.  Such a classical wavepacket should be related to a suitable coherent state in the
boundary theory.  Inserting this expression into (\ref{ads/cft}), taking one functional derivative and setting the source to zero gives a
contribution to the CFT 1-point function which goes as
\begin{equation}
\langle \phi_{wp} | O({\bf b}) | \phi_{wp} \rangle \sim \int d^3 x \sqrt{-g}  g^{\mu \nu} \partial _\mu \phi_{wp} (x) \partial _\nu K
(x,\bf{b}) . \labell{1ptfcn}
\end{equation}
In view of the extended Penrose diagram Fig.~\ref{fig:pen} with multiple asymptotic regions and horizons, the key question is what regions
are included in the integration over the bulk and which boundary components contribute.

The obvious application of the correspondence as stated in \cite{Witten:1998qj,Gubser:1998bc} would suggest that we should integrate in the
bulk over the entire spacetime, including the infinitely repeating regions in Fig.~\ref{fig:pen} (for relevant discussions see
\cite{Hemming:2002kd}).   The spacetime contains an infinite number of disconnected boundaries, so the dual CFT would presumably contain
infinitely many disconnected components.     However, boundary conditions would also have to be  specified  at the naked timelike
singularities beyond each inner horizon.  This is not very palatable.  In any case, the instability that we have found at the inner horizon
is expected to close off the spacetime there and thus limit the range of integration to include only the regions shown in
Fig.~\ref{fig:horizon},

In \cite{kos,leviross} prescriptions for computing the vacuum correlation functions  were given by appealing to analytic continuation from
the Euclidean section.  One prescription involves integrating the bulk point, $x$, over all of regions $1_{+\pm}$ and $2_{+\pm}$ shown in
Fig.\ \ref{fig:horizon}, up to, but not including the full Cauchy horizon.   This can be simply expressed by using the Poincare disc
coordinates or the outer Kruskal coordinates ($U_+$,$ V_+$) that were described earlier, since these coordinates cover same range, namely
all of  Fig.\ \ref{fig:horizon} except precisely the inner horizon.

In an alternative prescription, derived by a different analytic continuation of the same Euclidean section, we integrate over only the
regions outside the event horizon ($1_{+\pm}$) but over a contour in the complex $t$ plane which we shall describe below. In effect we use
BTZ coordinates in the outer region and exploit the excursions in the complex plane described earlier to move from $1_{++}$ to $1_{+-}$. It
was proposed in \cite{kos,leviross} that the parts of the contour that move purely in the imaginary time direction should be thought of as
encoding the physics behind the horizon.

In the case of the vacuum correlation functions,  both of these prescriptions are different analytic continuations of the same Euclidean
integral, and therefore they give the same result. It was therefore claimed in \cite{kos,leviross} that the integral over only the external
regions (plus the imaginary time parts of the integration contour)  encode information from behind the horizon.  However, in the vacuum
situation studied in these papers nothing special happens behind the horizon, so a skeptic might have read the result in the opposite
direction: the AdS/CFT prescription that integrates behind a horizon is really only sensitive to information that is accessible to the
asymptotic observer.   Our ability to include a wavepacket that is regular at the outer horizon and breaks down at the inner horizon
provides an interesting opportunity to test how a dramatic phenomenon localized behind the horizon is detected by the AdS/CFT
correspondence. Of course, when a state is present, we cannot perform a continuation to the Euclidean section; so it is not possible to
derive a prescription for computing amplitudes via such analytic continuations.  Instead we will adopt the prescriptions derived by
\cite{kos,leviross} in the vacuum since they should not change much if a small perturbation is introduced in the spacetime.   Of course, our
perturbation has dramatic effects at the inner horizon and we are precisely interested in the resulting breakdown of the semiclassical
approximation of the AdS/CFT correspondence.

\subsection{General considerations}
We are interested in seeing some sign of the breakdown of the semiclassical AdS/CFT correspondence due to the inner horizon instability. The
first step is to demonstrate that the contribution to CFT correlators from the vicinity of the outer horizon is finite and well defined.
Then we will study the contribution from the inner horizon.    We will be principally interested in the 1-point function given by the
integral (\ref{1ptfcn}).   Potential breakdowns in this integral can arise from the measure factor $\sqrt{-g}$ and from the coordinate
invariant piece $g^{\mu\nu} \partial_\mu\phi_{wp} \partial_\nu K$.  Here we will study the behavior of the integrand at the outer and inner
horizons and in the next section we will study the full integral.

\vspace{0.2in}

\noindent {\bf (i) The outer horizon:} Let us examine the contribution to the 1-point function in (\ref{1ptfcn}) from the vicinity of the
outer horizon in the region $1_{++}$. (The qualitative behavior will be similar in $1_{+-}$.)  The expression  for the bulk-boundary
propagator can be obtained from the $AdS_3$ bulk-boundary propagator by a sum over images. When both the boundary point and the bulk point
are contained in region $1_{++}$ it is given by \cite{Keski-Vakkuri:1998nw,Hemming:2002kd,leviross,kos}
\begin{equation}
K^{1_{++}1_{++}} (x,b') = c \sum_{l=-\infty}^{\infty} \biggl\{ \sqrt{u} \cosh\Bigl( \frac{r_-}{\Lambda ^2} \Delta t - \frac{r_+}{\Lambda}
\Delta \bar{\phi}_l \Bigr) -\sqrt{u-1} \cosh\Bigl( \frac{r_+}{\Lambda ^2} \Delta t - \frac{r_-}{\Lambda} \Delta \bar{\phi}_l \Bigr)
\biggr\}^{-2h_+} , \labell{bbprop}
\end{equation}
where $c$ is a constant, $\Delta t = t-t'$ and $\Delta \bar{\phi}_l = \bar{\phi} - \bar{\phi}' +2 \pi l$.   The spacetime integral giving
the CFT 1-point function (\ref{1ptfcn}) involves a measure factor $\sqrt{-g}$ and a coordinate invariant expression of the form $g^{\mu\nu}
\partial_\mu\phi_{wp} \partial_\nu K$.

This invariant piece can be evaluated near the event horizon, giving (ignoring for now the sum over $l$ in the propagator (\ref{bbprop}) and
switching to the $\phi_+ =\bar{\phi} -\Omega_H t$ coordinate)
\begin{eqnarray} \label{full eh ads expression}
g^{\mu \nu} \partial _\mu \phi_{wp} (x) \partial _\nu K (x,b')  \approx -2h_+ \{ \ldots \}^{-2h_+-1} e^{in\phi_+}  \frac{r_+}{r_+^2-r_-^2}
 \times \hspace{2.25in}\nonumber \\
\Biggl(  e^{-\kappa_+ r_*}   \biggl\{ f' e^{i\Omega_H n U} \biggl[ \exp \Bigl(\kappa_+ \Delta t - \frac{r_- \Delta \phi_+}{\Lambda} \Bigr)
\biggr] - g' e^{i\Omega_H n V} \exp \biggl[-\Bigl(\kappa_+ \Delta t - \frac{r_- \Delta \phi_+}{\Lambda} \Bigr) \biggr] \biggr\}
\nonumber \\
+ e^{-\kappa_+ r_*} i \Omega_H n  \biggl\{ f e^{i\Omega_H n U} \biggl[ \exp \Bigl( \kappa_+ \Delta t - \frac{r_- \Delta \phi_+}{\Lambda}
\Bigr)  \biggr] - g e^{i\Omega_H n V}  \exp \biggl[-\Bigl( \kappa_+ \Delta t - \frac{r_- \Delta \phi_+}{\Lambda} \Bigr)  \biggr] \biggr\}
 \nonumber \\
+  \biggl[ e^{i\Omega_H n U} (-f' -i \Omega_H n f) + e^{i\Omega_H n V } (g' +i\Omega_H n g) \biggr] \cosh
\Bigl(\frac{r_+ \Delta \phi_+}{\Lambda} \Bigr)  \nonumber \\
+\sinh \Bigl(  \frac{r_+ \Delta \phi_+}{\Lambda} \Bigr)  {in \over \Lambda} (f e^{i\Omega_H n U} +g e^{i\Omega_H n V}) \Biggr) , \ \ \ \ \ \
\end{eqnarray}
where
\begin{equation}
\{ \ldots \} = \sqrt{u} \cosh \Bigl( \frac{r_+}{\Lambda} \Delta \phi_+ \Bigr) - \sqrt{u-1} \cosh \Bigl( \kappa_+ \Delta t -
\frac{r_-}{\Lambda} \Delta \phi_+ \Bigr) . \labell{dots}
\end{equation}
Each $\Delta\phi_+$ in these expressions should really be a $\Delta \phi_{+,l}$ with the $2\pi l$ shifts coming from the propagator.  We are
neglecting the sum over $l$ because this does not contribute to any divergent behavior at the outer horizon.

It is easiest to examine the behavior of the integral (\ref{1ptfcn}) near the event horizon in the outer Kruskal coordinates ($U_+$, $V_+$,
$\phi_+$) defined in (\ref{nullouterkruskal}), as the coordinate system is well defined on the event horizon and it is easier to perform the
integrals. The metric becomes
\begin{eqnarray}
ds^2 &=& \Omega^2(r) dU_+dV_+ +r^2 (N^{\phi_+} dt +d \phi_+)^2 , \\
\Omega^2(r) &=& \frac{(r^2-r_-^2)(r+r_+)^2}{\kappa_+^2r^2 \Lambda^2} \Bigl( \frac{r-r_-}{r+r_-} \Bigr)^{r_-/r_+} , \\
N^{\phi_+} &=& \frac{r_-}{\Lambda r_+ r^2} (r^2-r_+^2) ,
\end{eqnarray}
where it is understood that $t$ and $r$ are now implicit functions of $U_+$ and $V_+$.   In these coordinates $\sqrt{-g}$ is finite along
the whole event horizon. In addition, one can check that  $\{\ldots\}$ (\ref{dots}) is finite along both the past and future event horizons.
Therefore, any possible divergent behavior must come from the terms in (\ref{full eh ads expression}).

Examining (\ref{full eh ads expression}) and using  (\ref{kruskal v transform}) we see that the only possibly divergent terms have the
general form of
\begin{equation} \label{f on eh}
f (\textrm{or} \ f') e^{i\Omega_H n U} e^{\kappa_+ \Delta t - \frac{r_- \Delta \phi_+}{\Lambda}} e^{-\kappa_+ r_*} \sim \frac{f(U)}{U_+} ,
\end{equation}
or
\begin{equation} \label{g on eh}
g (\textrm{or} \ g') e^{i\Omega_H n V} e^{-\kappa_+ \Delta t + \frac{r_- \Delta \phi_+}{\Lambda}} e^{-\kappa_+ r_*} \sim \frac{g(V)}{V_+} .
\end{equation}
(We can treat the derivative terms the same way, because we are only interested in the exponential part of $f$ and $g$.) On the past event
horizon the outgoing piece (\ref{f on eh}) is finite. To check the ingoing piece (\ref{g on eh}), recall that $g(V) \sim e^{\kappa_+ V}$ on
the past event horizon. We find
\begin{equation}
\frac{g(V)}{V_+} \sim \frac{e^{\kappa_+ V}}{V_+} \sim  1 ,
\end{equation}
and so is finite. A similar computation shows both terms to be finite on the future event horizon as well. Thus, the integrand is finite
along the entire event horizon and the integral for the 1-point function does not receive any divergent contributions from the event
horizon.

\vspace{0.2in}

\noindent {\bf (ii) The inner horizon:} We can carry out a similar set of calculations near the Cauchy horizon. When the bulk point is the
region $2_{++}$ we can find the bulk-boundary propagator by an analytic continuation \cite{Hemming:2002kd,kos,leviross}
\begin{equation}
K^{2_{++} 1_{++}} (x,b') = c' \sum_{l=-\infty}^{l=\infty} \biggl\{ \sqrt{u} \cosh\Bigl( \frac{r_-}{\Lambda^2}\Delta t - \frac{r_+}{\Lambda}
\Delta \bar{\phi}_l \Bigr) - \sqrt{1-u} \sinh\Bigl( \frac{r_+}{\Lambda^2}\Delta t - \frac{r_-}{\Lambda} \Delta \bar{\phi}_l \Bigr)
\biggr\}^{-2h_+} .
\end{equation}
The coordinate invariant part of the integrand in (\ref{1ptfcn}), i.e., $g^{\mu\nu} \partial_\mu\phi_{wp} \partial_\nu K$, evaluated near
the Cauchy horizon is (again ignoring the sum in the propagator for now and switching to $\phi_-$ defined in (\ref{phiminus})~)
\begin{eqnarray} \label{full ch ads expression}
g^{\mu \nu} \partial _\mu \phi_{wp} (x) \partial _\nu K (x,b') \approx -2h_+ \{ \ldots \}^{-2h_+-1} e^{in \phi_-} \frac{r_-}{r_+^2-r_-^2}
\times \hspace{1.9in} \nonumber \\ \Biggl(
 e^{\kappa_- \tilde{r}_*}
\biggl[ -\tilde{f}' e^{i\Omega_C n \tilde{U}} \exp\Bigl( -\kappa_- \Delta t - \frac{r_+}{\Lambda} \Delta \phi_- \Bigr) +\tilde{g}'
e^{i\Omega_C n \tilde{V}} \exp\Bigl(  \kappa_- \Delta t + \frac{r_+}{\Lambda} \Delta \phi_- \Bigr) \biggr] \nonumber
\\
+ e^{\kappa_- \tilde{r}_*} i \Omega_C n  \biggl[ - \tilde{f} e^{i\Omega_C n \tilde{U}}\exp\Bigl(  -\kappa_- \Delta t - \frac{r_+}{\Lambda}
\Delta \phi_- \Bigr)+\tilde{g} e^{i\Omega_C n \tilde{V}}\exp\Bigl( \kappa_- \Delta t + \frac{r_+}{\Lambda} \Delta \phi_-
 \Bigr) \biggr] \nonumber \\
- [ e^{i\Omega_C n \tilde{U}} (-\tilde{f}'-i\Omega_C n\tilde{f} \Bigr) +e^{i\Omega_C n \tilde{V}} (\tilde{g}'+i\Omega_C n\tilde{g} \Bigr)
\sinh\Bigl( \frac{r_-}{\Lambda} \Delta \phi_- \Bigr)
\nonumber \\
- \frac{in}{\Lambda} (\tilde{f} e^{i\Omega_C n \tilde{U}}+\tilde{g} e^{i\Omega_C n \tilde{V}})\cosh\Bigl(\frac{r_-}{\Lambda} \Delta \phi_-
\Bigr) \Biggr) , \ \ \ \ \
\end{eqnarray}
with
\begin{equation}
\{\ldots \} = \sqrt{u} \cosh\Bigl( \kappa_- \Delta t + \frac{r_+}{\Lambda} \Delta \phi_- \Bigr) + \sqrt{1-u} \sinh\Bigl( \frac{r_-}{\Lambda}
\Delta \phi_- \Bigr) . \labell{chldots}
\end{equation}
It is easiest to examine the behavior of the 1-point function integral (\ref{1ptfcn}) near the inner horizon using the inner Kruskal
coordinates $(U_-,V_-,\phi_-)$ given in (\ref{phiminus},~\ref{nullinnerkruskal}). The metric is given by
\begin{equation}
ds^2 = \tilde{\Omega}^2(r) dU_-  dV_- +r^2 (N^{\phi_-} dt +d \phi_-)^2 ,
\end{equation}
\begin{equation}
\tilde{\Omega}^2(r) = \frac{(r_+^2-r^2)(r+r_-)^2}{\kappa_-^2r^2 \Lambda^2} \Bigl( \frac{r_+-r}{r_++r} \Bigr)^{r_+/r_-} ,
\end{equation}
\begin{equation}
N^{\phi_-} = \frac{r_+}{\Lambda r_- r^2} (r^2-r_-^2) ,
\end{equation}
and the determinant of the metric and the overall factor (\ref{chldots})$\{\ldots\}$ are finite on the inner horizon. Near the Cauchy
horizon we have
\begin{equation}
r-r_- \approx - 2 r_- \biggl( \frac{r_+-r_-}{r_++r_-} \biggr)^{r_+/r_-} U_- V_- , \labell{nearchrtransform}
\end{equation}
\begin{equation}
e^{\kappa_- t} = \sqrt{\frac{-U_-}{V_-}} , \labell{nearchttransform}
\end{equation}
(note that $U_-/V_- < 0$ in region $2_{++}$).

In a similar analysis to the one performed near the event horizon, we find the leading (potentially divergent) terms near the Cauchy horizon
have the form
\begin{equation} \label{f on ch}
\tilde{f} (\textrm{or} \ \tilde{f}') e^{i \Omega_C n \tilde{U}} e^{\kappa_- \tilde{r}_* } e^{-\kappa_- \Delta t - \frac{r_+}{\Lambda} \Delta
\phi_- } \sim \frac{\tilde{f}(\tilde{U})}{U_-} e^{\kappa_- t' -\frac{r_+}{\Lambda} \Delta \phi_-},
\end{equation}
or
\begin{equation} \label{g on ch}
\tilde{g} (\textrm{or} \ \tilde{g}') e^{i \Omega_C n \tilde{V}} e^{\kappa_- \tilde{r}_* } e^{\kappa_- \Delta t + \frac{r_+}{\Lambda} \Delta
\phi_- } \sim \frac{\tilde{g}(\tilde{V})}{V_-} e^{-\kappa_- t' +\frac{r_+}{\Lambda} \Delta \phi_-}.
\end{equation}
On the left Cauchy horizon ($U_-=0$) the leftmoving piece (\ref{g on ch}) is finite since we know that $\tilde{g}(\tilde{V})$ itself is well
behaved and $V_-$ is nonzero there. Using (\ref{past inner horizon behavior}), the rightmoving piece (\ref{f on ch}) goes as
\begin{eqnarray}
\frac{\tilde{f}(\tilde{U})}{U_-} &\sim& \frac{(Res(EH) + Res(DHe^{- \frac{\beta_*}{2} (\omega -n \Omega_H)})) |_{\omega=\omega_{2,0}} }{U_-}
 \nonumber \\
 &+& \frac{(Res(EH)+ Res(DHe^{- \frac{\beta_*}{2} (\omega -n \Omega_H)}))e^{\kappa_+ \tilde{U}} |_{\omega=\omega_{1,0}} }{U_-}
 \sim U_-^{-1} + U_-^{\kappa_+/\kappa_- -1} .
 \labell{leftchintegrand}
\end{eqnarray}
This diverges as $U_- \to 0$ at the horizon.  Similarly, on the right Cauchy horizon ($V_-=0$) the rightmoving piece (\ref{f on ch}) is
finite, while the leftmoving piece (\ref{g on ch}) diverges as
\begin{eqnarray}
\frac{\tilde{g}(\tilde{V})}{V_-} &\sim& \frac{[Res(FH)+Res(CHe^{- \frac{\beta_*}{2} (\omega -n \Omega_H)}] |_{\omega_{2,0}} }{V_-} \nonumber \\
&+& \frac{\exp\Bigl[ -2h_+ \frac{ (r_+ +r_-)}{\Lambda^2}\tilde V \Bigr] Res(FH)  |_{\omega_{3,0}} }{V_-} \sim V_-^{-1} + V_-^{\frac{2h_+
r_-}{r_+-r_-}-1} ~~. \labell{rightchintegrand}
\end{eqnarray}

To summarize, we have seen that the leading behavior of the integrand in the AdS/CFT expression for the CFT 1-point function (\ref{1ptfcn})
is finite on the event horizon and divergent on the inner horizon.   To see whether this leads to a breakdown or divergence in the CFT
1-point function we have to examine the full integral in (\ref{1ptfcn}) and understand how it behaves in the different prescriptions for
integrating over the the BTZ spacetime that were discussed above and in \cite{kos,leviross}.

\subsection{Computing the 1-point function}

As discussed before there are different prescriptions for carrying out the 1-point function integral in (\ref{1ptfcn}).   The obvious
procedure would have been to integrate over the entire spacetime, or at least over all of the regions in Fig.~\ref{fig:horizon} including
the Cauchy horizon.  As we discussed, becaused of the timelike singularities beyond the Cauchy horizon, this is potentially problematic. The
second procedure, from \cite{leviross}, is derived by analytic continuation in the vacuum case and requires an integral up to, but not
including the Cauchy horizon.  This can be done either in the disc coordinates or in the outer Kruskal coordinates that were described
before.   Finally, a third prescription, also derived in \cite{leviross} in the vacuum case via analytic continuation, integrates up to and
including the outer horizon in both asymptotic regions and over part of the complexified BTZ coordinate plane.  Our goal will be to show how
all of these semiclassical procedures break down because of the presence of the wavepacket in our setup and the resulting instability of the
inner horizon.

\vspace{0.2in}

\noindent {\bf (i)~Integrating through the Cauchy horizon: }   Suppose that the correct region of integration includes the Cauchy horizon in
$2_{++}$ and in $2_{+-}$.  This could either mean that we integrate over the entire spacetime (as one would naively suppose) or that we
integrate over all of the regions in Fig.~\ref{fig:horizon} including the Cauchy horizon. Near the left Cauchy horizon, the leading behavior
of the integral is
\begin{equation}
\int d^3 x \sqrt{-g}  g^{\mu \nu} \partial _\mu \phi_{wp} (x) \partial _\nu K (x,{\bf b}) \sim \int dV_- \int d \phi_- \int^0  dU_- \{
\ldots \} ^{-2h_+-1} U_-^{-1} ,
\end{equation}
where the lower limit is unimportant for our considerations. Using our expression for $\{ \ldots \}$ (\ref{chldots}) and the transformation
equations (\ref{nearchrtransform}, \ref{nearchttransform}), we find
\begin{eqnarray}
&&\int dV_- \int d \phi_- \int^0  dU_- \{ \ldots \} ^{-2h_+-1} U_-^{-1} \nonumber \\
&\sim& \int dV_- \int d \phi_- \int^0  dU_- \frac{1}{U_-\Bigl[U_- e^{-\kappa_- t' -\frac{r_+}{\Lambda} \Delta \phi_- }  + V_- e^{\kappa_- t'
+\frac{r_+}{\Lambda} \Delta \phi_- }  +\sinh\Bigl( \frac{r_-}{\Lambda} \Delta \phi_- \Bigr) \Bigr]^{2h_+ +1}} \nonumber \\
&\sim& \int dV_- \int d \phi_- \ln (-U_-) |^0 \sim \infty , \labell{leftchintegral}
\end{eqnarray}
where we have only kept the most divergent piece. A similar calculation shows that the integral diverges logarithmically in $V_-$ near the
right Cauchy horizon as well.

In \cite{kos} a potential divergence was found coming from the region near the non-rotating BTZ singularity.  This arose because of the sum
on images in the bulk-boundary propagator.  The situation here is very different.  It was already shown in \cite{leviross} that the sum on
images does not lead to divergences in correlation functions from regions near the Cauchy horizon.   The same applies here.  We are
uncovering a different phenomenon associated with the focusing of the wavepacket at the inner horizon.

In \cite{kos} the potential divergence arising from the sum on images could be regulated by an $i\epsilon$ prescription derived from
Euclidean continuation of the bulk-boundary propagator.  In addition there were cancellations between some contributions to the correlation
function integrals at the past and future BTZ singularities.   Such regularizations and cancellations cannot occur here. First, let us
consider whether divergences from the right and left Cauchy horizons could cancel, and second, whether divergences from the Cauchy horizon
in $2_{++}$ could cancel those in $2_{+-}$.  From (\ref{f on ch}, \ref{g on ch}) we see that the two pieces have different dependences on
the boundary (primed) coordinates. Therefore, for general values of $t', \phi' _-$ the contributions from the right and left Cauchy horizons
are different and the two terms cannot cancel.  Turning to the other possibility it is easy to show that the wavepacket is not generically
symemtric between the $2_{++}$ and $2_{+-}$  regions and therefore there are no cancellations.

Since the divergences will not cancel we can attempt to regulate them.  One might try to use a Cauchy Principal Value prescription for
integrating through the inner horizon. However, this regularization prescription will fail for higher derivative terms that we have not
explicitly included in the present analysis. The other alternative is to choose an $i \epsilon$ prescription to define the 1-point function
integrals (\ref{1ptfcn}) since a similar prescription rendered integrals in the non-rotating vacuum case finite \cite{kos}. However, because
of the absence of a Euclidean continuation in our case,  and the fact that our wavepackets are complex as well as containing both left and
rightmoving ($U_-$ and $V_-$  dependent) pieces,  there is no obvious principled  approach to regulating our integrals in this way without
introducing pinched singularities.

In addition, stringy ($\alpha'$) corrections to the supergravity action will contain terms with more and higher derivatives. These terms
will lead to even worse divergences in the integral.   The essential lesson here is that backreaction and strong curvature effects will be
important near the inner horizon and we should expect to need the full string theory, not just the semiclassical supergravity approximation.
Since the dual CFT is finite and unitary, the properly computed 1-point function must be finite. Thus, we interpret the divergence of the
integral (\ref{1ptfcn}) as a breakdown in the semiclassical approximation to the AdS/CFT correspondence.   In this prescription it is clear
that correlation functions are sensitive to the physics near the inner horizon.

\vspace{0.2in} \noindent {\bf (ii)~The prescription in disc coordinates: } In \cite{leviross} a prescription for computing AdS/CFT
correlators was found (via analytic continuation from the Euclidean section) that integrates over the regions shown in
Fig.~\ref{fig:horizon}. This is conveniently expressed in Poincare disc coordinates (see \cite{leviross} for the details of this coordinate
system) which cover the same regions as the outer Kruskal coordinates $(U_+,V_+,\phi_+)$. These coordinates cover all of $1_{+\pm}$ and all
of $2_{+\pm}$ except for the Cauchy horizon.   In other words, the range for the BTZ radial coordinate is  $r_- < r < \infty$.   (This
differs from the the analysis above in that the actual inner horizon is excluded.  Since this is a set of measure zero it can only matter if
the integrand in (\ref{1ptfcn}) diverges there.)

Above, we saw that the integral contributing to the 1-point function picked up a divergence from the contribution on the Cauchy horizon.
Here we must omit the actual horizon, but include points that are infinitesimally separated from it.
%The prescription of \cite{leviross} integrates up to an infinitesimal distance, $\epsilon$, from the inner
%horizon.
>From (\ref{leftchintegral}) the resulting behavior near the left Cauchy horizon is
\begin{equation}
\int dV_- \int d \phi_- \int^{-\epsilon}  dU_- \{ \ldots \} ^{-2h_+-1} U_-^{-1} \sim \ln (-U_-) |^{-\epsilon} \sim \ln \epsilon
\end{equation}
where we cut off the Cauchy horizon by limiting the integral to the range $-U_- \geq \epsilon$. This integral is  again ill-defined as
$\epsilon \to 0$.   In addition, stringy effects and higher derivative interactions will produce terms that go like $1/\epsilon ,
1/\epsilon^2 , \ldots$ which will not be suppressed relative to the semiclassical supergravity approximation. As above, this integral is
sensitive to the instability at the inner horizon and the semiclassical approximation to the AdS/CFT correspondence breaks down.  Also as
discussed above, these divergences do not cancel between different horizons and cannot be regulated away in a straightforward manner.

\vspace{0.2in} \noindent {\bf (iii)~The prescription in BTZ coordinates: } In the other prescription given in \cite{kos,leviross} we are
instructed to integrate using BTZ coordinates only over the region outside the event horizon, but over a contour in the complex $t$ plane.
Since we are only integrating over the exterior region, the way in which the correlator can see the instability inside the horizon is more
subtle.

\begin{figure}
\begin{center}
    \includegraphics[width=0.9\textwidth]{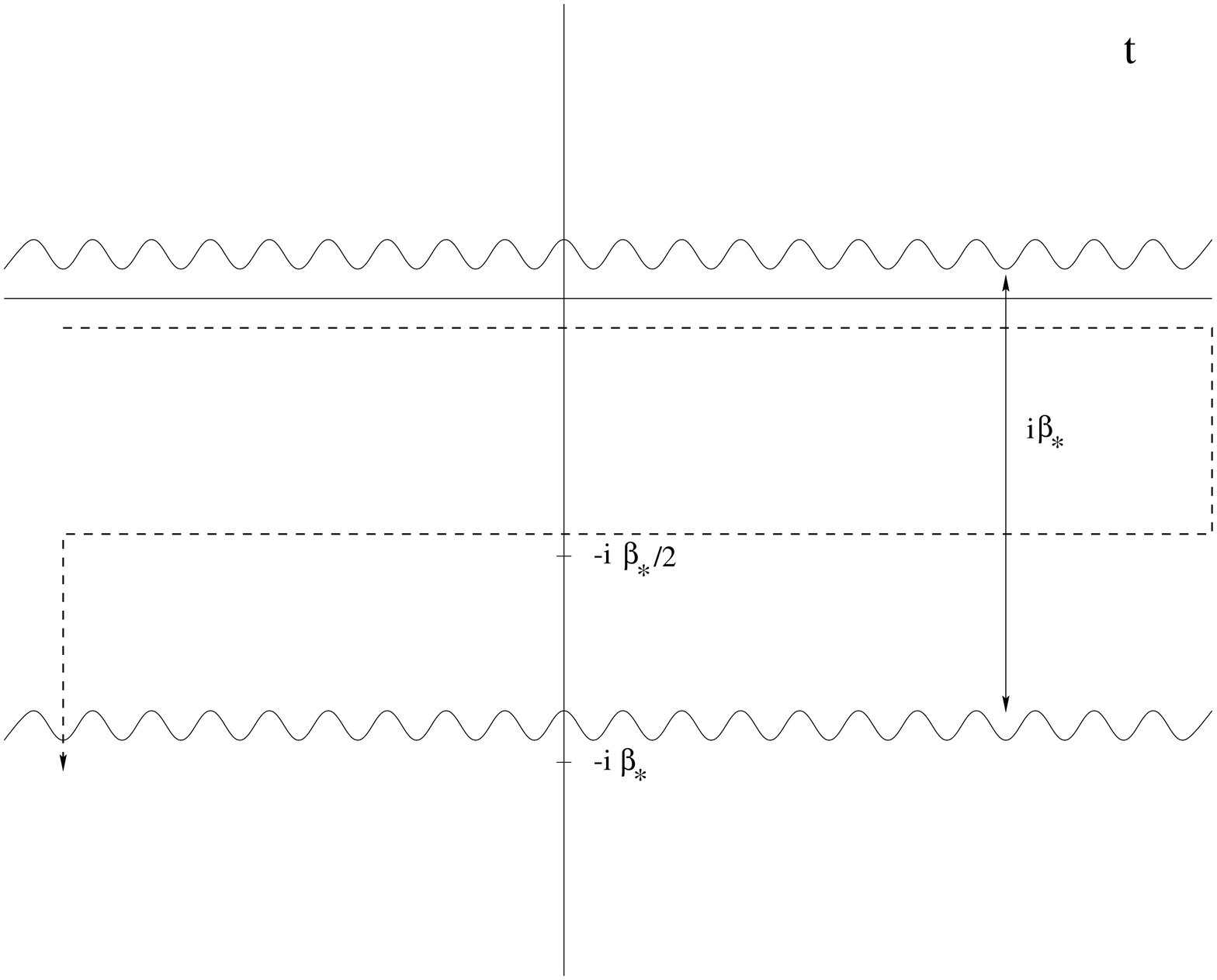}
\end{center}
\caption{The complex $t$ contour for integration in BTZ coordinates. The wavy lines are the branch cuts that restrict the range of imaginary
time. While choosing different locations for the cuts can move these lines, the contour will always either cross or hit the cuts. }
\label{fig:btzcontour}
\end{figure}

In this prescription (\ref{1ptfcn}) becomes
\begin{equation}
\langle \phi_{wp} | O({\bf b}) | \phi_{wp} \rangle \sim \int _C d^3 x \sqrt {|g|} \,  g^{\mu \nu}  \, \partial_\mu K_C(x,{\bf b}) \,
\partial_\nu \phi_{wp,C} (x) ,
\end{equation}
where the $t$ integral is carried out over the contour $C$ shown in Fig.\ \ref{fig:btzcontour}, and we integrate only over the asymptotic
regions $1_{+\pm}$ down to the event horizon. Both $\phi_{wp}$ and $K$ are defined over the complex values of $t$. For the bulk-boundary
propagator, this analytic continuation is straightforward and takes us over the different possible choices of bulk and boundary points, as
well as the Euclidean bulk-boundary propagator. However, a problem arises when we consider the  presence of $\phi_{wp}$ in the analytically
continued integral. Recall, due to the presence of the branch cuts in Kruskal coordinates, we found a restriction on the imaginary values of
BTZ time, $t$, given by (\ref{t restriction}). However, the integration over the vertical sections of the contour shown in Fig.\
\ref{fig:btzcontour} violates this restriction and hence crosses the branch cut in general.  As a result the integral is not well defined as
is since we will be integrating over the same spacetime point with different values for the wavepacket $\phi_{wp}$.   By shifting the cuts
the best we can do is to have the both the upper horizontal section and the lower end point of the contour just touch the cuts. While we
could try to define the integral anyway via a suitable infinitesimal deformation of the contour, it is not clear exactly what deformation to
use and what its physical justification would be.

Notice that it is precisely the presence of the vertical sections of the contour,  representing the entanglement between the dual CFTs
living on the two BTZ boundaries, which lead to the breakdown of the integral. The horizontal sections can be well defined within the
restrictions due to the branch cuts.  Since we understand the breakdown as arising from an instability at the inner horizon it is tempting
to conclude that  the presence of an entanglement between the two boundary CFTs can encode information from behind the horizon of the black
hole.

\subsection{Remarks}

We have seen that the inner horizon instability is picked up  when we are integrating over the region including the Cauchy horizon as a
divergence arising ultimately from the focusing effect. When integrating only over the outer region, the required contour leads to an
integral that is not defined.   Of course, if we were to do the 1-point function computation in the boundary theory itself, we must get a
finite, well-defined answer. Thus, we interpret the problems with the bulk prescription as a  breakdown of the semiclassical approximation
to the AdS/CFT correspondence  and as a signal that backreaction must be accounted for.    It would be natural, for example, for the
integration contour in Fig.~\ref{fig:btzcontour} to be modified by backreaction.  For example, since the temperature of the black hole
determines the lower endpoint of the contour in the complex time plane, in the context where we have thrown in an additional wavepacket one
might ask whether the correct contour involves the temperature seen by an asymptotic observer before or after the wavepacket has entered the
black hole.  More generally, the dynamical process of equilibration should be involved.   In this regard, previous studies of backreaction
and its effects on black hole radiance (see, e.g., \cite{backreact}) may be interesting.

We have mostly focused on the physics of the inner horizon of rotating black holes.  In fact, a similar set of calculations can be carried
out for the non-rotating BTZ black hole and establishes that a normalizable wavepacket has large backreaction at the singularity. The
analysis in BTZ coordinates is similar to the above and we find that the contour crosses the cuts. For the prescription in Kruskal
coordinates a general derivative interaction will possess two types of divergences, those coming from the sum over images in the
bulk-boundary propagator at the singularity and those coming from the blueshifting of the wavepacket. In \cite{kos} a regularization of the
divergences coming from the bulk-boundary propagator was found through analytic continuation from the Euclidean section that rendered
integrals finite. This regularization amounts to adding a small, imaginary piece, $i \epsilon$ to the radial coordinate $r$ and then taking
the limit $\epsilon \to 0$ at the end of the calculation. Since a normalizable wavepacket is not regular in the Euclidean section, it is
difficult to unambiguously define a regularization for its contribution to CFT correlators. Without regularization we find that higher
derivative interactions diverge and stringy corrections are no longer suppressed. If we try to use the regularization from \cite{kos}, we
find that since the wavepacket is in general complex, any terms that contain the wavepacket and its complex conjugate will have opposite $i
\epsilon$ prescriptions and pinched singularities seem to occur. It is important to emphasize that the rotating case is fundamentally
different. Without the wavepacket, there are no divergences associated with the inner horizon and so no regularization is necessary
\cite{leviross}, in contrast to the singularity in the non-rotating case. Therefore, in the rotating case, there is no suggestion at all
from the Euclidean section as to how to regulate the divergences that are present.

One should interpret AdS/CFT calculations in the BTZ background as computing  CFT expectation values in a state that entangles  the two
otherwise disconnected boundary CFT components \cite{Horowitz:1998xk,vijay2,maldacena,kos}
\begin{equation} \label{entangled state}
|\Psi \rangle = \frac{1}{\sqrt{Z}} \sum_n e^{ -\frac{\beta_* E_n}{2} -\frac{\mu \beta_* l_n}{2} } |E_n, l_n \rangle_1 \otimes |E_n, l_n
\rangle_2 ,
\end{equation}
where $Z$ is a normalization factor, $|E_n, l_n \rangle$ is an energy and angular momentum eigenstate in one of the CFTs and $\mu$ is the
chemical potential related to the angular velocity of the outer horizon of the black hole. The entanglement is between the CFTs living on
the two disconnected boundaries of the black hole (i.e. in $1_{++}$ and $1_{+-}$).   If we only insert operators in one CFT, we can trace
over states in the second CFT, and obtain a thermal expectation value for an $n$-point function $G_n$
\begin{equation}
G_n = \textrm{Tr} \{e^{-\beta_* H} T[{\cal O}(t_1),\ldots {\cal O}(t_n)] \} ,
\end{equation}
where we have suppressed the angular dependence. Writing out this trace in terms of evolution operators one can obtain the usual thermal
contour in \cite{kos}. In our case, we have perturbed away from thermal equilibrium and our state is no longer given by (\ref{entangled
state}). Tracing over the states in the second CFT leads to an expression of the form
\begin{equation}
G_n = \textrm{Tr} \{\rho T[{\cal O}(t_1),\ldots {\cal O}(t_n)] \} ,
\end{equation}
where $\rho$ is some non-equilibrium density matrix.  Notably it will include initial correlations. It is no longer possible in general to
obtain the usual thermal contour \cite{non-eq}, which is precisely what we expected. In principle, one could carry out this computation in
the boundary theory and so hope to obtain information about the resolution of the inner horizon instability (and the resulting singularity)
using holography.

\section{Conclusions and discussion} \label{sec-con}
We showed that the addition of normalizable modes in the rotating BTZ black hole opens up a wealth of new physics. At the classical level,
we  found it necessary to construct wavepackets to regularize behavior on the outer horizon. Even with this regularization a focusing
instability is always present at the inner horizon and so it is likely that a singularity forms there. Continuity, single-valuedness and
positive energy of wavepackets as we move between black hole regions control the location of branch cuts in the complexified coordinate
planes.  We showed that AdS/CFT correlators are sensitive to the inner horizon instability and that the prescription that integrates only
outside the black hole learns about the black hole interior through these cuts.   This result reinforces observations in
\cite{kos,leviross,Fidkowski:2003nf, Kaplan:2004qe,Balasubramanian:2003kq} that point to the importance of the complexified coordinate plane
as a repository of physical information in time-dependent backgrounds.

In this regard, it is interesting that the boundary of the Euclidean black hole is a single, connected torus, and that one can also view
motion in the imaginary time direction as connecting the two asymptotic regions and boundaries in the Lorentzian section. Thus, the vertical
parts of the contour in Fig.~\ref{fig:btzcontour} correspond to moving along this torus from one boundary CFT to the other.    It is worth
noting that the analytic continuation in \cite{kos,leviross} that led to this contour prescription is not in fact unique.  For example, for
the purpose of computing CFT correlation functions in the Hartle-Hawking vacuum for the BTZ black hole, one could have used a contour in
Fig.~\ref{fig:btzcontour} in which the lower horizontal section was at $Im t=-i \beta_*/3$, or even a contour that has a diagonal piece. For
vacuum correlators many such contours would give the same CFT correlation functions as computed from the AdS/CFT prescription in
\cite{kos,leviross}.   However, the interpretation of these other contours both in bulk terms and in CFT terms is not clear.  (Also see
\cite{Herzog:2002pc} for similar remarks.)   It would be interesting to either eliminate or understand the physical interpretation of these
other contours.

In principle, one could directly compute CFT correlators in the presence of the non-equilibrium density matrix corresponding to an infalling
wavepacket.   These are expected to be finite and well defined, and  should carry information concerning how string theory resolves the
inner horizon instability of rotating AdS black holes.

\begin{acknowledgments}
We would like to thank V.\ Jejjala, D.\ Mateos, A.\ Naqvi, and J.\ Simon for extensive discussions. We thank S.\ Ross for valuable
discussions and collaboration during the early stages of this work. V.B.\ thanks Aruna Beatrix Balasubramanian who was born during the
writing of this paper and assisted in many an editing session. Work on this project was supported by the DOE under cooperative research
agreement DE-FG02-95ER40893, by the NSF under grant PHY-0331728 and by an NSF Focused Research Grant DMS0139799 for ``The Geometry of
Superstrings".
\end{acknowledgments}


\end{document}